\documentclass[%
reprint,
superscriptaddress,
amsmath,amssymb,
prb,
]{revtex4-1}

\usepackage{graphicx}
\usepackage{dcolumn}
\usepackage{bm}
\usepackage{color}
\usepackage{xfrac}
\usepackage[font=small,labelfont=bf,justification=justified,format=plain]{caption}
\usepackage{subcaption}

\begin{document}

\preprint{APS/123-QED}

\title{Evidence for a pressure-induced spin transition in olivine-type LiFePO$_{4}$ triphylite}

\author{Maribel N\'u\~nez Valdez}
\email{mari\_nv@gfz-potsdam.de}
\affiliation{GFZ German Research Centre for Geosciences, Telegrafenberg, 14473 Potsdam, Germany}

\author{Ilias Efthimiopoulos}
\email{iliefthi@gfz-potsdam.de}
\affiliation{GFZ German Research Centre for Geosciences, Telegrafenberg, 14473 Potsdam, Germany}
\affiliation{Institute of Earth and Environmental Science, University of Potsdam, Karl-Liebknecht-Strasse 24-25, 14476 Potsdam-Golm, Germany}%

\author{Michail Taran}
\affiliation{Natl Acad Sci Ukraine, Inst Geochem Mineral and Ore Format, UA-03680 Kiev 142, Ukraine}%

\author{Jan M\"uller}
\affiliation{GFZ German Research Centre for Geosciences, Telegrafenberg, 14473 Potsdam, Germany}
\affiliation{Institute of Geology and Mineralogy, University of Cologne, Zülpicher Str. 49b, 50674 Cologne, Germany}%

\author{Elena Bykova}
\affiliation{FS-PE, PETRA III, Deutsches Elektronen Synchrotron, 22607 Hamburg, Germany}%

\author{Monika Koch-M\"uller}
\affiliation{GFZ German Research Centre for Geosciences, Telegrafenberg, 14473 Potsdam, Germany}

\author{Max Wilke}
\affiliation{Institute of Earth and Environmental Science, University of Potsdam, Karl-Liebknecht-Strasse 24-25, 14476 Potsdam-Golm, Germany}%
\date{\today}
\begin{abstract}
We present a combination of first-principles and experimental results regarding the structural and magnetic properties of olivine-type LiFePO$_4$ under pressure. Our investigations indicate that the starting $Pbnm$ phase of LiFePO$_4$ persists up to 70~GPa. Further compression leads to an isostructural transition in the pressure range of ~70-75 GPa, inconsistent with a former theoretical study. Considering our first-principles prediction for a high-spin to low-spin transition of Fe$^{2+}$ close to 72 GPa, we attribute the experimentally observed isostructural transition to a change on the spin state of Fe$^{2+}$ in LiFePO$_4$. Compared to relevant Fe-bearing minerals, LiFePO$_4$ exhibits the largest onset pressure for a pressure-induced spin state transition.
\end{abstract}
\pacs{Valid PACS appear here}
\maketitle
\section{Introduction}
Compounds belonging to the lithiophilite-triphylite series Li$M2$PO$_4$ ($M2$~=~Mn,~Fe), are commonly found throughout the Earth's crust and mantle in pegmatites with high Li and P contents\cite{Losey2004}. At ambient conditions, they are isostructural with (Mg,Fe)$_2$SiO$_4$--olivine (space group, SG, 62, $Pbnm$ and $Z=4$), one of the most abundant minerals ($\sim$50-60\% vol) in the upper-mantle of our planet ($<$410~km depth and pressure $P$~$<$13.5~GPa)\cite{Ringwood1975}. Fig. 1 shows this olivine-type structure, which can be seen as a distorted hexagonal close-packing of oxygen anions with corner--sharing $M2$O$_6$, and edge-sharing LiO$_6$ octahedra aligned in parallel chains along the $\mathbf{b}-$axis. The P ions are in PO$_4$ tetrahedra linking the octahedral layers.

The transition metal $M2$ in these Li$M2$(PO$_4$) phosphate-materials is octahedrally coordinated (Fig. 1), which leads to crystal field splitting ($\Delta$) of the $M2$ 3$d$ electrons into three lower--energy levels, $t_{2g}$, and two higher--energy levels, $e_g$. At low pressures, Hund's rule predicts that, for example, $M2$=Fe$^{2+}$ (ferrous iron) adopts the high-spin (HS) state with $S=2$ ($t_{2g}^{4}e_{g}^{2}$). However, as pressure increases, $\Delta$ increases at the expense of the spin-pairing exchange energy $J$ ($J<\Delta$), and Fe$^{2+}$ undergoes a spin transition to the energetically more favorable low-spin (LS) state with $S=0$ ($t_{2g}^{6}e_{g}^{0}$). Using computational and experimental approaches, this pressure-induced spin crossover has been investigated and reported in several Fe-bearing minerals of the Earth's lower-mantle (in the range of 660--2890 km depth and 23--135 GPa) such as (Mg,Fe)O--ferropericlase\cite{Badro2003,Lin2005,Speziale2005,Tsuchiya2006,Goncharov2006,Lin2007,Crowhurst2008,Wentzcovitch2009,Marquardt2009,Antonangeli2011,Wu2013,Holmstrom2015,Wu2014,Hsu2010}, (Mg,Fe)SiO$_3$--bridgmanite\cite{Badro2004,Li2004,McCammon2008,Lin2008,Hsu2010-pv,Hsu2011,Potapkin2013,Hsu2014}, and (Mg,Fe)CO$_3$--magnesiosiderite\cite{Mattila2007,Lavina2009,Lavina2010,Nagai2010,Farfan2012,Lin2012,Liu2014,Lobanov2015,Hsu2016,Muller2016,Muller2017}. The study of this HS-to-LS transition is important as it has effects on the structural, optical, elastic, and thermodynamic properties of the Fe-bearing host minerals, and in consequence potential and profound geophysical repercussions in the interpretation of seismic velocities and geodynamical modeling\cite{Marquardt2009,Lin2013,Wu2014}.
\begin{figure}[h]
\begin{subfigure}{0.480\textwidth}
  \includegraphics[width=\textwidth]{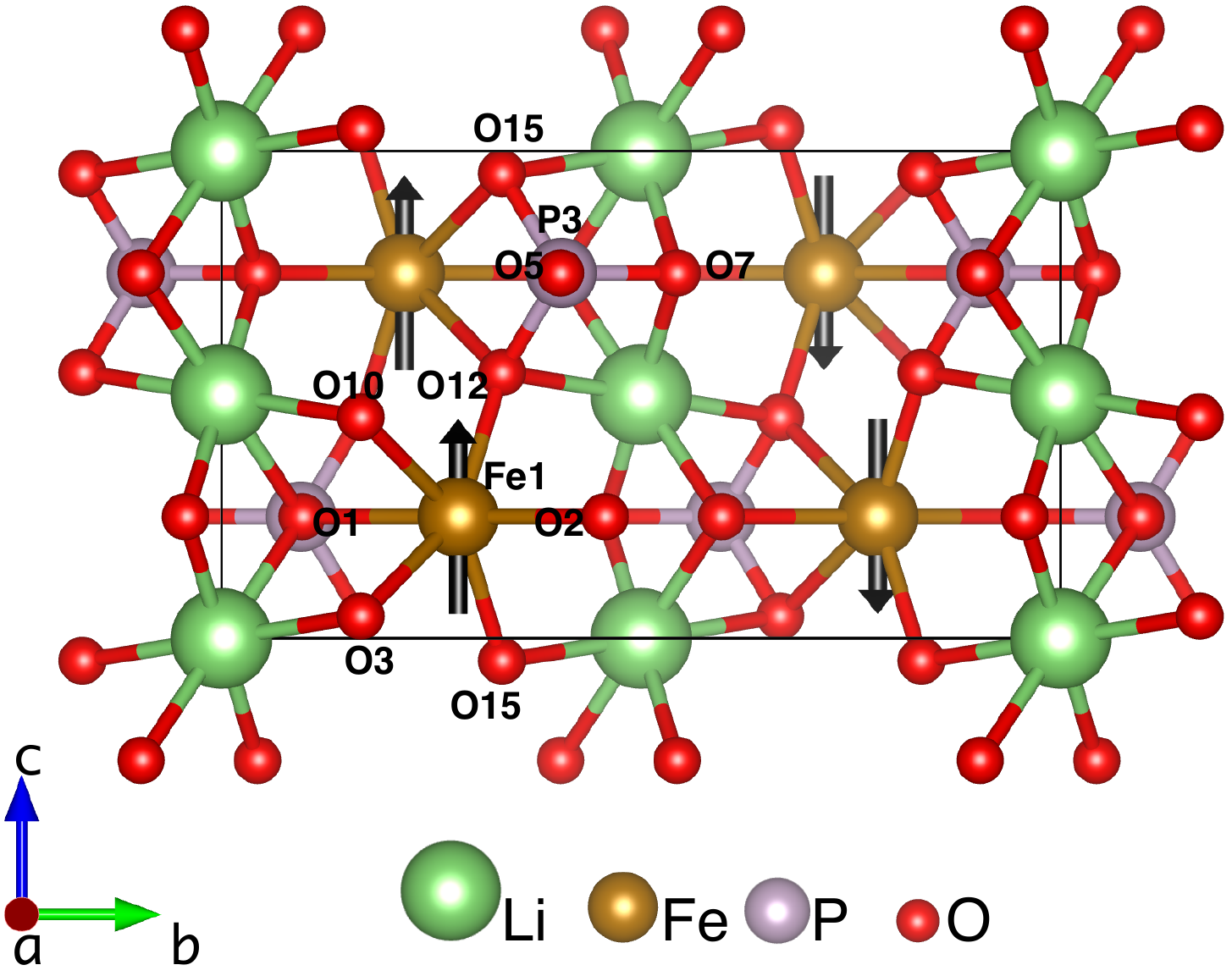}
  \caption{Ground state of $Pbnm$--LiFePO$_4$ . The AFM ordering is indicated with black arrows.}
  \label{metastable_a}
\end{subfigure}
\begin{subfigure}{0.30\textwidth}
        \includegraphics[width=\textwidth]{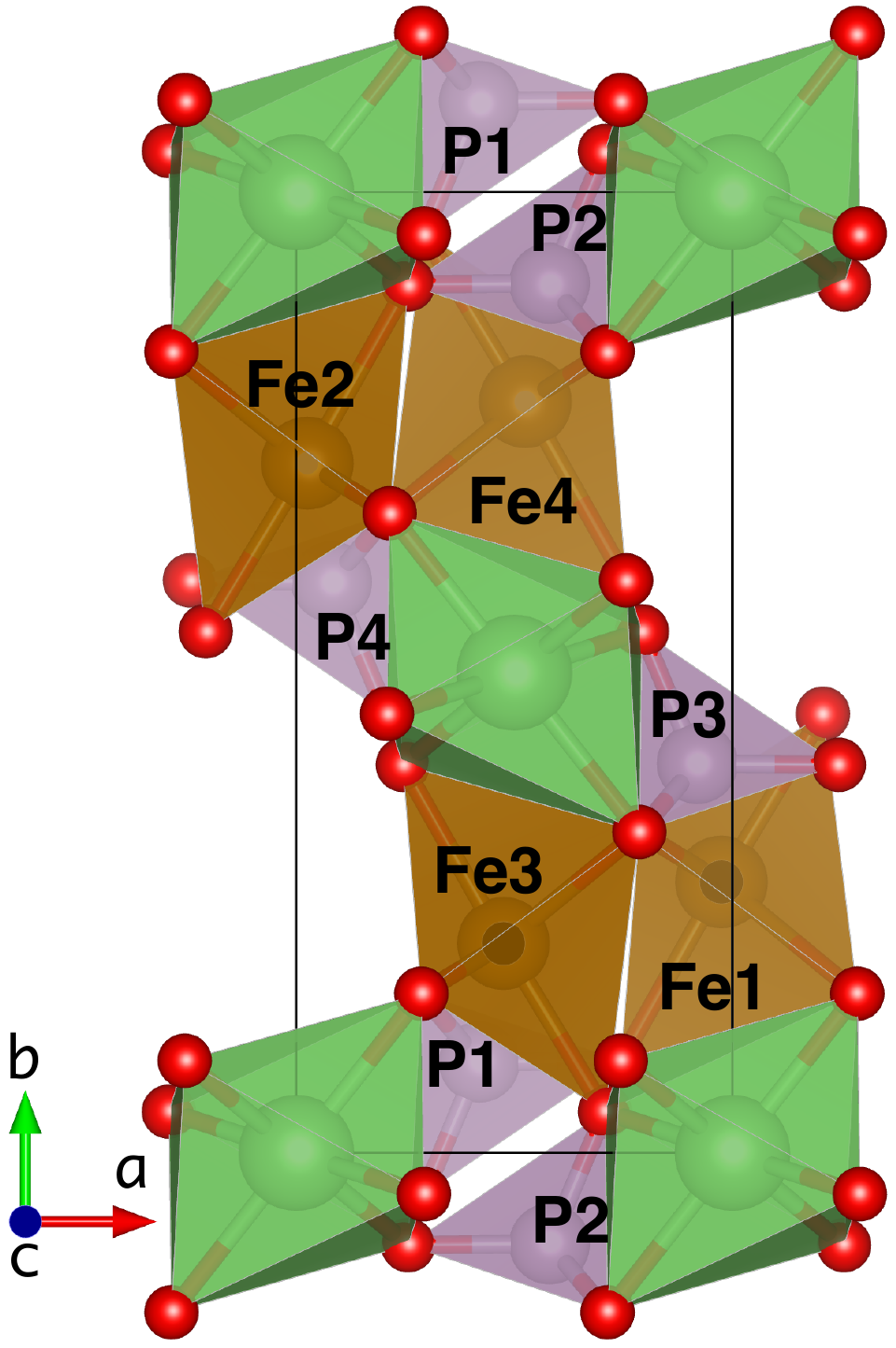}
        \caption{Polyhedral representation of LiFePO$_4$.}
        \label{metastable_b}
    \end{subfigure}
\caption{(Color online) At ambient pressure, LiFePO$_4$-triphylite ($M2=$Fe) crystallizes in the olivine-type structure (SG 62, $Pbnm$, $Z=4$) with AFM ordering. Selected Fe, P and O atoms are indicated for later analysis in Sec. IV. (Crystal structures in this paper are visualized using VESTA\cite{vesta}.)}
\label{metastable_ab}
\end{figure}
In addition to the aforementioned minerals, it was reported, using density functional theory (DFT)\cite{dft1,dft2}, that LiFePO$_4$-triphylite also undergoes a HS-to-LS spin crossover at about 52~GPa, although, not in the $Pbnm$ structure but rather as a $Cmcm$ high-pressure (HP) and high-temperature (HT) phase, which was shown to become the ground state at $\sim$4 GPa \cite{Lin2011}. Experimentally, the $Pbnm\rightarrow Cmcm$ structural transformation was observed at 6.5~GPa and 1173~K \cite{Garcia2001}. Recently, however, some controversy regarding the validity of the structural and spin transitions has emerged, as recent experimental and computational studies addressing the structural stability of LiFePO$_4$ under pressure\cite{Dodd2007,Dong2017} have shown that the $Pbnm$ phase is stable up to $\sim$30~GPa. Thus, leaving somewhat open and unresolved the issues of whether or not 1) the $Pbnm\rightarrow Cmcm$ structural phase transition occurs in triphylite at room temperature (RT), and if not, 2) does the $Pbnm$ phase undergo a HS-to-LS transition similar to other Fe-bearing minerals\cite{Badro2003,Badro2004,Mattila2007}?

Aside of the geophysical relevance of these Li$M2$PO$_4$ olivine-type materials, in recent years, LiFePO$_4$--triphylite has also attracted considerable interest as a storage cathode material in rechargeable Li-batteries (for a review see e.g. Ref.~[\onlinecite{Wang2015}] and references therein). Thus, a thorough study of triphylite will allow us to better understand the properties of this material and its potential high-pressure applications.

Partly motivated by a possible pressure-induced HS$\rightarrow$LS transition in LiFePO$_4$, we have expanded the investigated high-pressure structural and vibrational behavior of triphylite up to 80~GPa at RT, by means of synchrotron-based X-ray diffraction (XRD), and  infrared (IR) and Raman spectroscopy. Additionally, we have performed \textit{ab initio} calculations to investigate the structural stability of LiFePO$_4$-triphylite under pressure, and the potential spin crossover in the energetically most favorable phase. By comparing the experimentally determined volume as a function of pressure to the theoretical trends throughout the pressure range considered (0--90~GPa), we provide evidence for a pressure-induced HS-to-LS transition in the $Pbnm$--LiFePO$_4$ phase. After presenting our experimental (Sec. II A) and computational methods (Sec. II B), we give comprehensive results concerning experimental determination of vibrational and structural parameters (Secs. III A and III B), and computational structural and enthalpy trends (Sec. III C) as a function of pressure. We then discuss the spin crossover in LiFePO$_4$ (Sec. IV) and finally present our conclusions (Sec. V). Characterization of samples, additional figures and tables with more experimental measurements and computational results can be found in the \textbf{Supplemental Material}.
\section{Methods}
\subsection{Experimental details}
We have investigated two synthetic LiFePO$_{4}$ samples: the first one was commercial powder (Sigma-Aldrich 97$\%$, sample A), and the second sample was $\mu$m-sized single crystals (sample B) synthesized by combined high-pressure and high-temperature treatment of the aforementioned LiFePO$_4$ powder in the multi-anvil press installed at GFZ\cite{Koch-MullerMugnaioliRhedeEtAl2009} as follows (MA 502): sample A was enclosed in a Pt capsule, whereas the 18/11 assembly was used. Pressure was raised to 1.4 GPa and then temperature to 1373 K (melting) and the sample remained there for 30 minutes. Within 2 hours, temperature was slowly decreased down to 973 K. The multi-anvil press was rotated for 2 hours at this temperature. Throughout the whole experiment the pressure was kept at 1.4 GPa.

The characterization of sample A by means of XRD and Raman spectroscopy at ambient conditions revealed the presence of two additional impurity phases at a $\sim$30\% concentration (as estimated by the relative Bragg peak intensities), i.e., the rhombohedral and monoclinic modifications of Li$_{3}$Fe$_{2}$(PO$_{4}$)$_{3}$, both of them being Fe$^{3+}$--bearing and common during the preparation of LiFePO$_{4}$\cite{JulienZaghibMaugerEtAl2012}. In addition, M\"{o}ssbauer spectroscopy evidenced a 30~$\%$ content of Fe$^{3+}$ for sample A, in excellent agreement with our XRD measurements. More information on the characterization of the starting samples are provided in the \textbf{Supplemental Material}.

Gasketed diamond anvil cells (DACs) equipped with low-fluorescence type II-as diamonds of 250~$\mu$m and 200~$\mu$m culet diameters were used for pressure generation. The rhenium gaskets were preindented to $\sim$25~$\mu$m thickness, with holes of 100~$\mu$m diameters acting as sample chambers in separate runs. Argon served as pressure transmitting medium (PTM) in all spectroscopic experiments, whereas neon was used for the XRD measurements. The ruby fluorescence method was employed for pressure calibration\cite{Syassen2008}.

The high-pressure mid-IR (MIR) measurements at RT were conducted on LiFePO$_{4}$ powder pressed into a thin film (sample A), with the Vertex 80v FTIR spectrometer coupled to a Hyperion~2000 microscope at GFZ, within the 500--1800~cm$^{-1}$ spectral range. We used a KBr beamsplitter and a MCT Hg-Cd-Te detector. The spectra were averaged over 512~scans with a spatial resolution of 2~cm$^{-1}$. For details of the thin film preparation for IR experiments see e.g. Ref.~[\onlinecite{MroskoKoch-MuellerSchade2011}].

Our high-pressure RT Raman measurements were conducted with a Horiba Jobin Yvon LabRam HR800 VIS spectrometer at GFZ, equipped with a blue ($\lambda$~=~473 nm) solid state laser operating at a power of 2~mW outside the DAC. The same LiFePO$_{4}$ thin film measured with MIR (sample A), as well as single crystals of sample B with typical dimensions of 40~$\times$~40~$\mu$m$^{2}$ were measured within the 100--1200~cm$^{-1}$ frequency region.

The angle-resolved high-pressure XRD measurements were performed at the Extreme Conditions Beamline P02.2 of PETRA III (Hamburg, Germany)\cite{LiermannKonopkovaMorgenrothEtAl2015} with an incident X-ray wavelength $\lambda$~=~0.289~{\AA} and a beam size of 2~$\mu$m~$\times$~2~$\mu$m. Two-dimensional XRD patterns were collected with a fast flat panel detector XRD1621 from PerkinElmer (2048 pixels $\times$ 2048 pixels, 200~$\times$~200~$μ$pixel size) and processed with the FIT2D software\cite{Fit2D}. Refinements were performed using the GSAS+EXPGUI software packages\cite{Toby2001}. Due to the presence of impurities in sample A as we discussed earlier, we investigated only sample B with XRD by crushing the small single crystals into powder between the diamonds of the DAC.
\subsection{Computational details}
The computational results presented in our work were obtained by applying density functional theory (DFT)\cite{dft1,dft2} within the projector-augmented plane wave (PAW)\cite{blochl,kresse99} method as implemented in the VASP (ver. 5.4.4) software\cite{kresse96a, kresse96b,kresse99} installed in the supercomputer JURECA\cite{jureca}. We employed primarily the general-gradient approximation (GGA) in the Perdew-Burke-Ernzerhof (PBE)\cite{perdew2} prescription for the exchange-correlation potential, and for comparison we also tested the local density approximation (LDA)\cite{LDA}. The default PAW potentials with the valence configurations of 1$s^2$2$p^2$ for lithium, 3$d^7$4$s^1$ for iron, 2$s^2$3$p^3$ for phosphorous, and 2$s^2$2$p^4$ for oxygen were used throughout our calculations. We adopted a plane-wave kinetic energy cutoff of 500~eV and a uniform $\Gamma$--centered grid with 2$\pi\times$0.032~\AA$^{-1}$ spacing for reciprocal space sampling of our 28-atom cells for pressures between $-5$ to 100~GPa. The Hubbard $U$ correction\cite{Anisimov1991,Dudarev1998} was added to the localized Fe~3$d$ electrons (LDA+U and PBE+U), and we assessed its value in the range 0--5.3~eV, but found $U=2.5$~eV to better agree with the experimentally determined lattice parameters and the pressure at which, evidence of a spin transition is observed in LiFePO$_4$ (see Fig. 4 in \textbf{Supplemental Material}). Convergence of our calculations at each pressure was assumed when the forces on each atom were smaller than 1~meV/\AA~ and the total energy changes less than $10^{-8}$~eV. Triphylite structures were set up at each target pressure within the $Pbnm$ and $Cmcm$ space groups and subsequently fully relaxed (internal coordinates and lattice parameters) using a conjugate gradient minimization approach. At zero pressure, the ordering of the spin of Fe$^{2+}$, i.e., antiferromagnetic (AFM) and ferromagnetic (FM) configurations, was also probed.
\section{Results}
\subsection{Vibrational spectroscopy under pressure}
\begin{figure}
\includegraphics[width=0.5\textwidth]{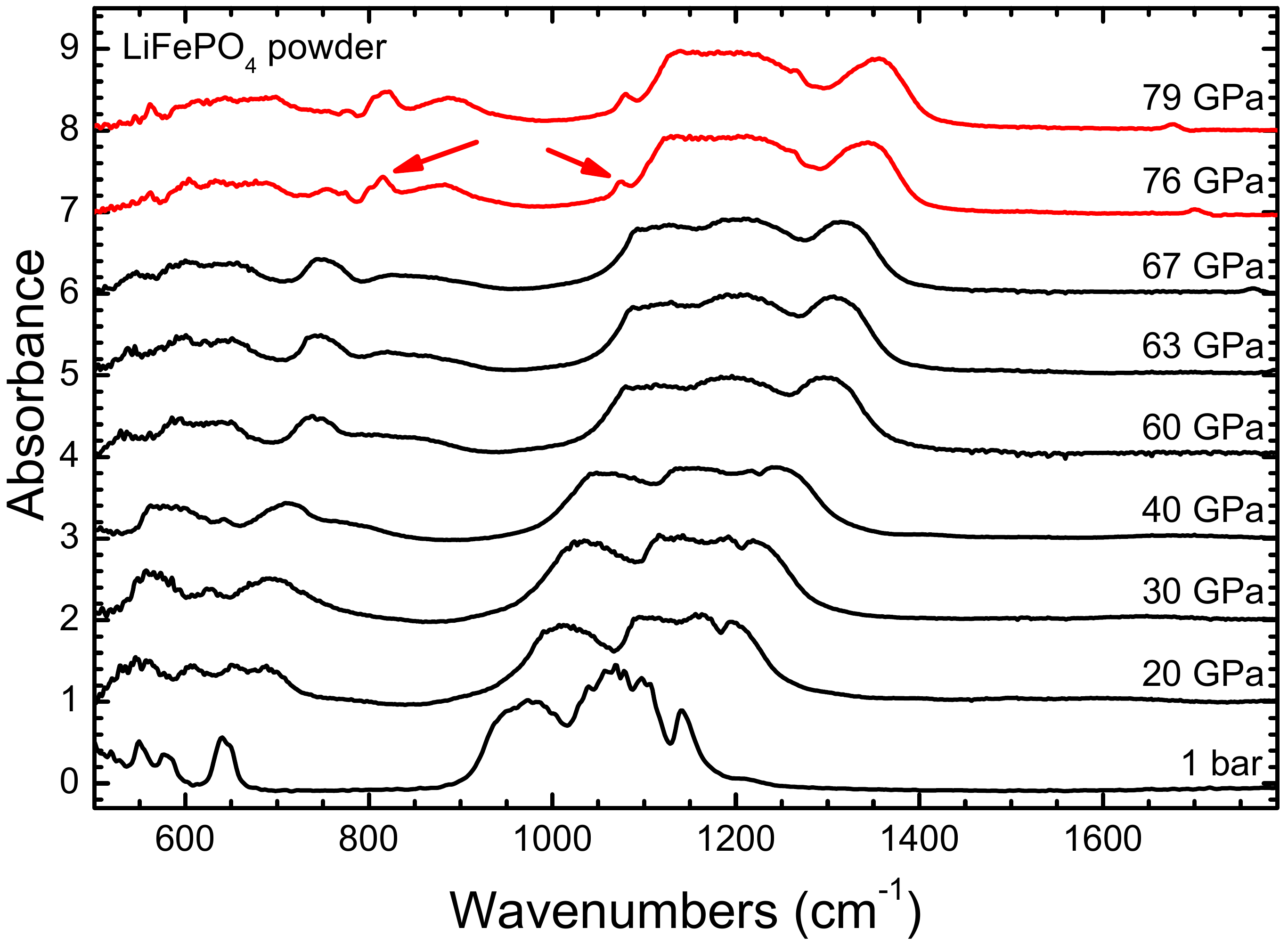}
\caption{\label{fig2} (Color online) MIR spectra of LiFePO$_{4}$ sample A at various pressures. The black and red spectra correspond to the $Pbnm$ and the HP-$Pbnm$ phases, respectively. The red arrows indicate the new MIR bands (see text).}
\end{figure}

\begin{figure*}
\begin{subfigure}{0.35\textwidth}
  \includegraphics[width=\textwidth]{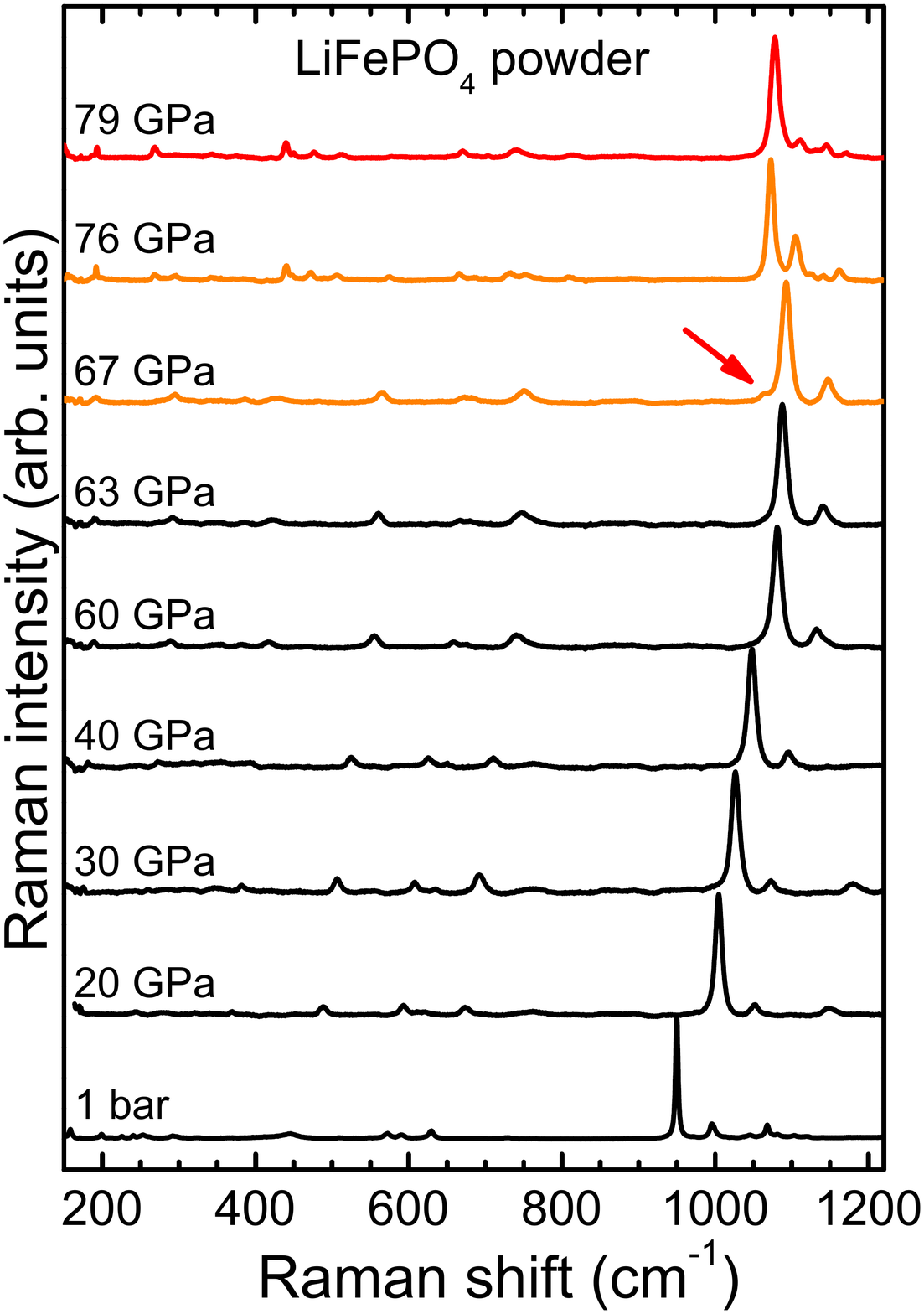}
  \caption{}
   \label{Raman_1}
\end{subfigure}
\begin{subfigure}{0.35\textwidth}
  \includegraphics[width=\textwidth]{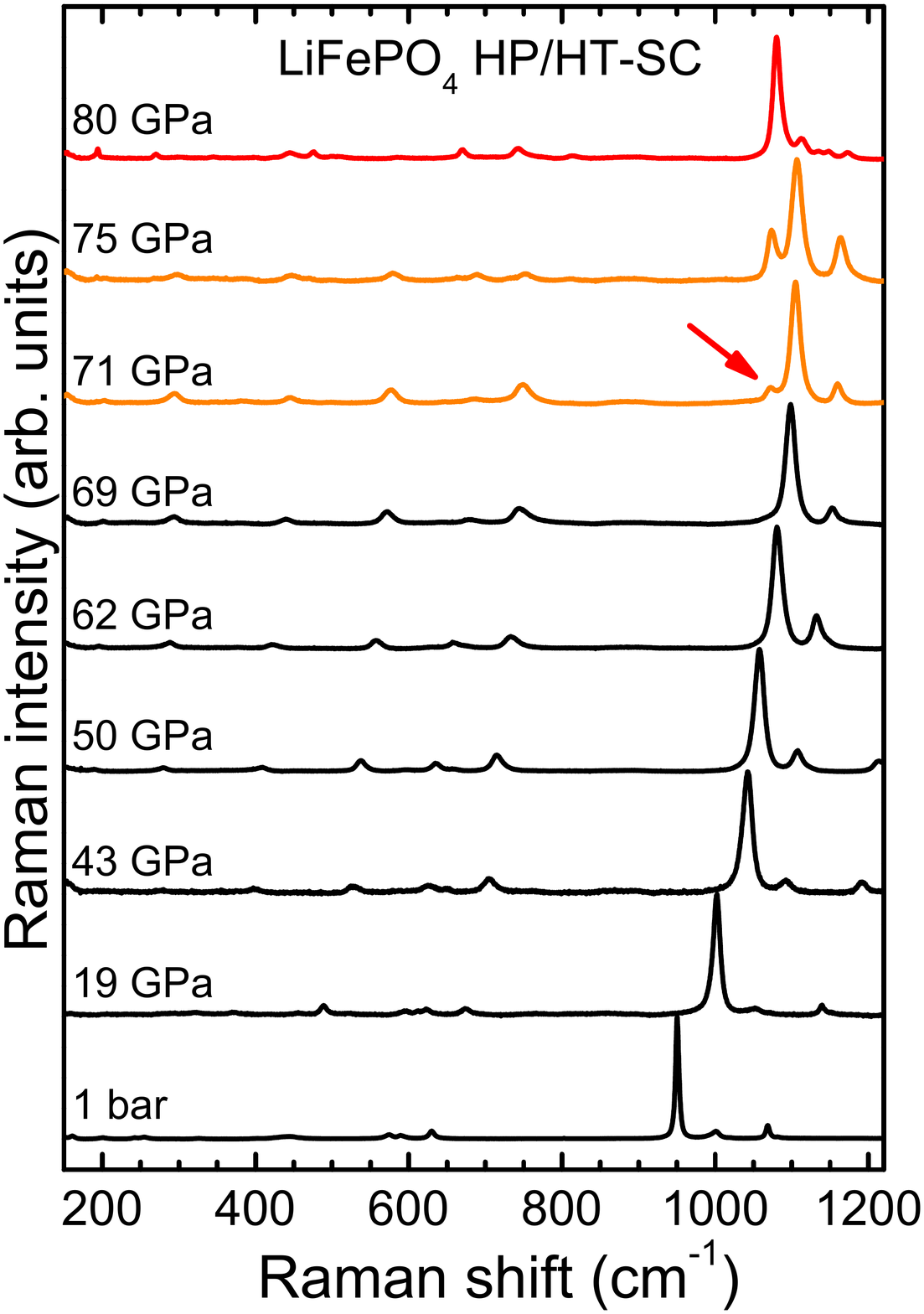}
  \caption{}
        \label{Raman_2}
\end{subfigure}
\caption{(Color online) Raman spectra of LiFePO$_{4}$ (a) sample A, and (b) sample B at selected pressures. The black, red, and orange spectra correspond to the $Pbnm$ and the HP-$Pbnm$ phases, and the coexistence range, respectively. The red arrows indicate the new Raman features in both cases (see text).}
\label{fig3}
\end{figure*}

\begin{figure}
\begin{subfigure}{0.35\textwidth}
  \includegraphics[width=\textwidth]{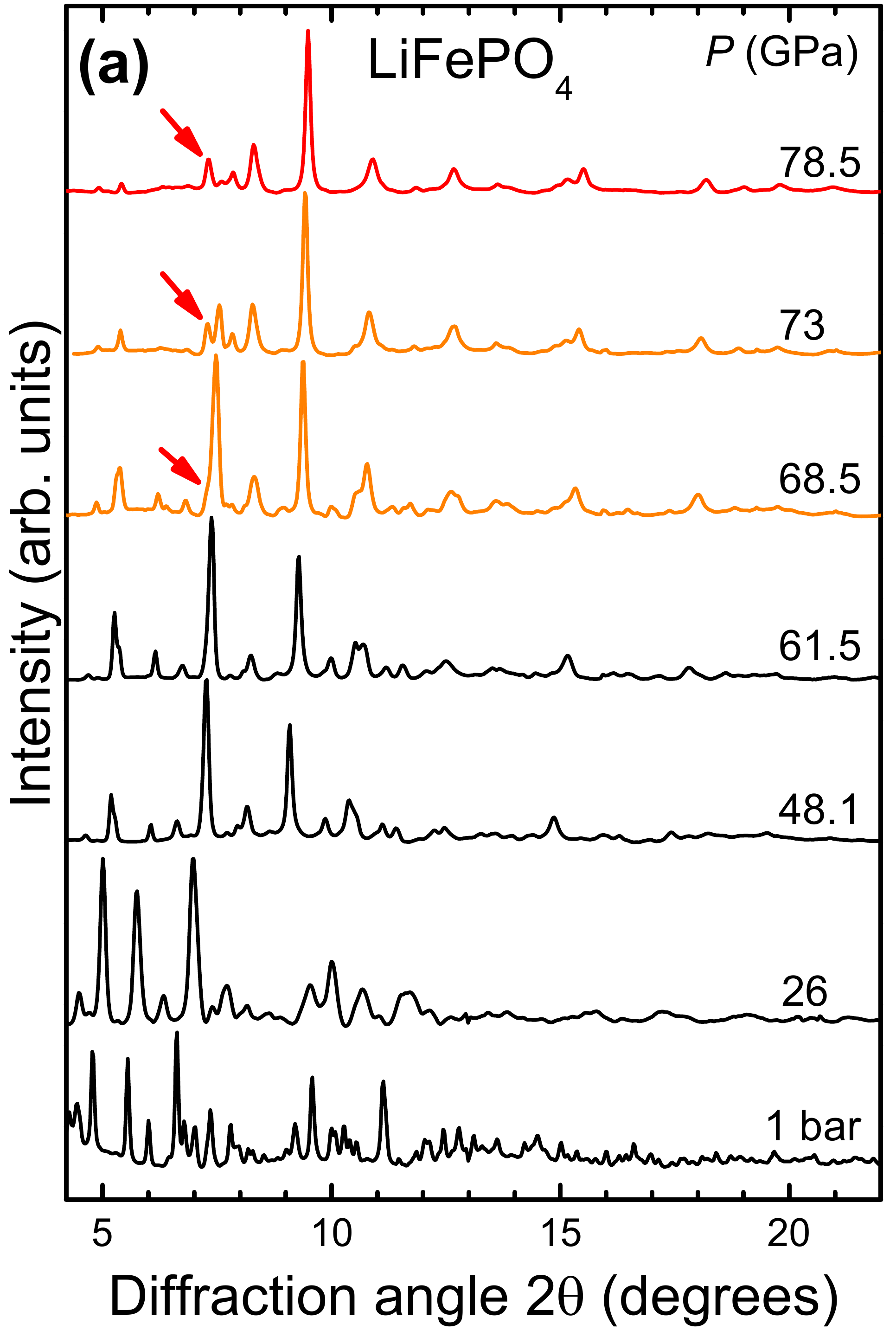}
   \label{XRD_1}
\end{subfigure}
\begin{subfigure}{0.35\textwidth}
  \includegraphics[width=\textwidth]{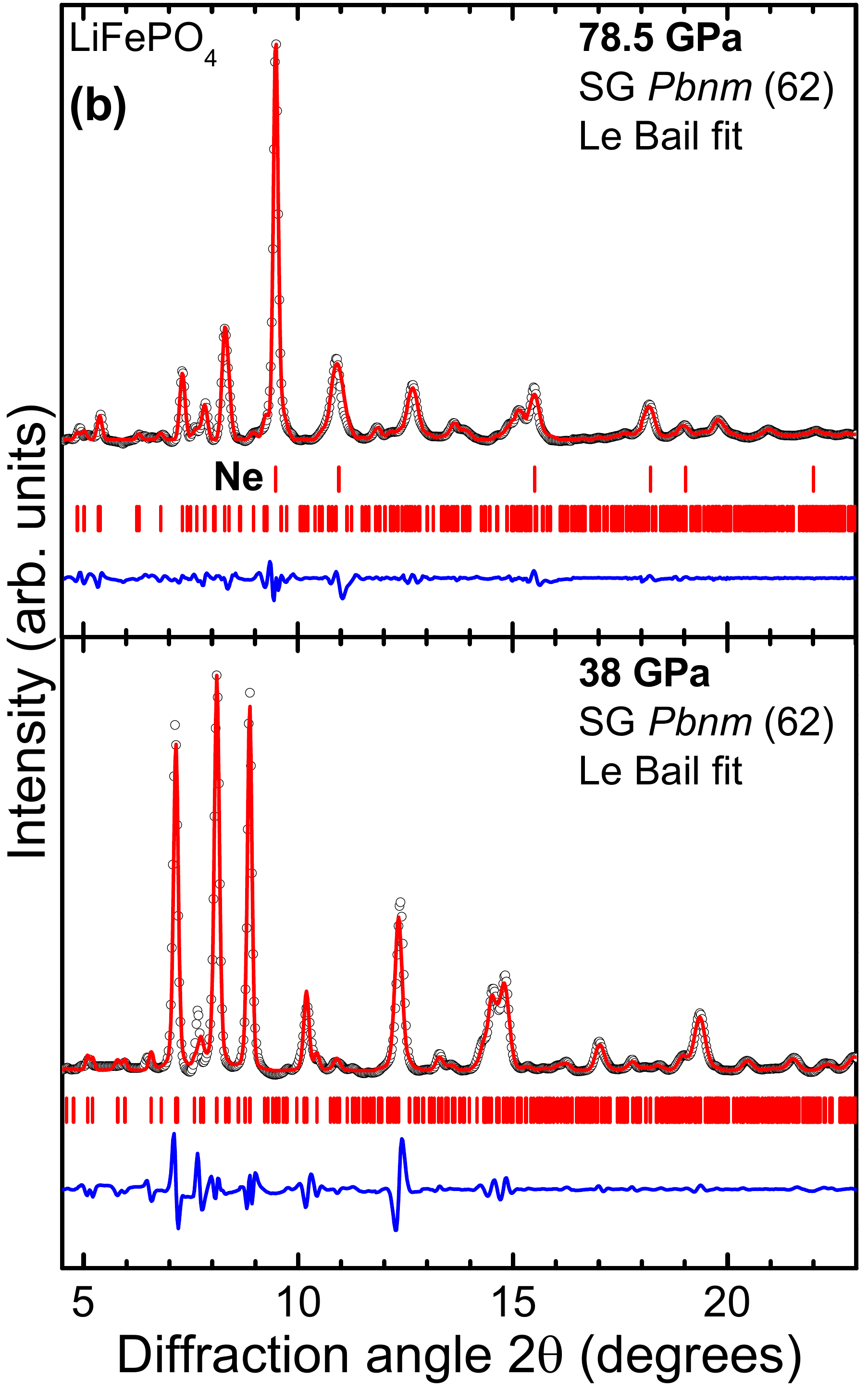}
        \label{XRD_2}
    \end{subfigure}
\caption{(Color online) (a) Selected XRD patterns of LiFePO$_{4}$ (sample B) at various pressures ($T=300$~K, $\lambda=0.29${\AA}). The $Pbnm$ phase is shown in black, the HP-$Pbnm$  phase in red, and their coexistence range is depicted in orange. The red arrows indicate the new XRD features (see text). (b) Examples of Le Bail refinements at 38~GPa (bottom) and 78.5~GPa (top). The black circles and the red solid lines correspond to the measured and the fitted patterns, whereas their difference is depicted as a blue line. Vertical ticks mark the Bragg peak positions for the $Pbnm$ phase and the neon PTM (only top).}
\end{figure}
According to group theory\cite{Paques-LedentTarte1974,Paques-LedentTarte1973,ParaguassuFreireLemosEtAl2005}, the expected IR- and Raman-active modes in LiFePO$_{4}$ are:
\begin{equation}
  \Gamma = 11A_{g} + 11B_{1g} + 7B_{2g} + 7B_{3g} + 14B_{1u} + 14B_{2u} + 10B_{3u}.
\end{equation}
Thus, LiFePO$_4$ has 29 Raman active modes ($11A_{g} + 11B_{1g} + 7B_{2g} + 7B_{3g}$) and 38 IR active modes ($14B_{1u} + 14B_{2u} + 10B_{3u}$). In Fig. 2, we present selected high-pressure MIR spectra collected for sample A. The ambient-pressure MIR spectrum of LiFePO$_{4}$ is consistent with the reported literature\cite{BurbaFrech2004,Ait-SalahDoddMaugerEtAl2006}. The intense broad MIR bands located between 1000--1200~cm$^{-1}$ correspond to internal antisymmetric P--O stretching vibrations $\nu_{3}$, the broad band centered at $\sim$960~cm$^{-1}$ belongs to the symmetric P--O stretching motion $\nu_{1}$, and the IR region between 400--800~cm$^{-1}$ is mainly attributed to the internal bending motions of the PO$_{4}$ tetrahedra, $\nu_{2}$ and $\nu_{4}$\cite{Ait-SalahDoddMaugerEtAl2006,BurbaFrech2004,Paques-LedentTarte1974}. The complete assignment can be found in Table 1 in the \textbf{Supplemental Material}.

Increasing pressure leads to the frequency upshift and a broadening of the recorded MIR bands, owing partly to the non-hydrostatic conditions imposed by the Ar PTM upon increasing pressure\cite{Klotz2009}, as well as the polycrystalline form of our sample. Nevertheless, close to 70~GPa we can detect the appearance of new IR bands between 700--800 cm$^{-1}$ and at 1000 cm$^{-1}$, indicating a pressure-induced transition of the $Pbnm$ LiFePO$_{4}$ structure. We will identify this high-pressure modification as HP-$Pbnm$ from now on. Interestingly, we can observe that the broad band at 750~cm$^{-1}$ is apparently vanishing in the new phase, with no other significant changes taking place throughout the transition.

Concurrently with our MIR experiment, we performed Raman spectroscopic investigations on the same LiFePO$_{4}$ thin film (sample A). Selected Raman spectra at various pressures are shown in Fig. 3(a). We note that the strongest Raman peak located at 950~cm$^{-1}$ corresponds to the symmetric P--O $\nu_{1}$ stretching vibration\cite{ParaguassuFreireLemosEtAl2005,Paques-LedentTarte1974}. Increasing pressure leads to the appearance of a small shoulder in the vicinity of the aforementioned $\nu_{1}$ mode at 67~GPa, indicating the phase transition of LiFePO$_{4}$ towards HP-$Pbnm$ at this pressure and in excellent agreement with our MIR results. Further compression results in the intensity enhancement of this shoulder at the expense of the 950~cm$^{-1}$ peak. Moreover, we can observe the appearance of several low-intensity Raman features (Fig. 2 in \textbf{Supplemental Material}).

In order to test whether this pressure-induced transition is possibly a byproduct of the polycrystalline nature of our sample or its impurities (see \textbf{Supplemental Material}), we have additionally conducted high-pressure Raman investigations on the HP/HT-synthesized LiFePO$_{4}$ single crystals. The relevant Raman spectra are plotted in Fig. 3(b). As we can observe clearly, the situation is identical with that of sample A, i.e., a new Raman peak in the vicinity of the strong 950~cm$^{-1}$ mode emerges at 71~GPa. We note that the Raman spectra at $\sim$80~GPa for both samples are practically identical, indicating that the pressure-induced $Pbnm\rightarrow$HP--$Pbnm$ structural transition of LiFePO$_{4}$ is inherent to the material and independent of the starting condition of the sample. At this stage, we cannot assign the high-pressure modification to the predicted non-magnetic $Cmcm$ phase (calculated to be adopted close to 52~GPa)\cite{Lin2011}, as its Raman spectrum has not been reported.
\subsection{Structure under pressure}
\begin{table*}%
\caption{\label{Table_1}%
Volume $V$, bulk modulus $B$, and its pressure derivative $B'$ for $Pbnm$-LiFePO$_{4}$ and its HP-$Pbnm$ phase, as obtained by a Birch-Murnaghan EoS\cite{Birch1947} fitted to our measured and computed $P$-$V$ data ($\diamond$). ``Fixed'' values were not allowed to vary during the fitting. Other results are from $\star$: Ref.~[\onlinecite{MaxischCeder2006}], $\dagger$: Ref. ~[\onlinecite{Dong2017}], $\ddagger$: Ref.~[\onlinecite{Dodd2007}]. }
\begin{ruledtabular}
\begin{tabular}{lldddd}
LiFePO$_4$ phase & Method & \multicolumn{1}{c}{\textrm{$P$ (GPa)}} &  \multicolumn{1}{c}{\textrm{$V$ ({\AA}$^{3}$)}} &   \multicolumn{1}{c}{\textrm{$B$ (GPa)}} &  \multicolumn{1}{c}{\textrm{$B'$ (GPa)}}\\
   \colrule
  $Pbnm$ & EXP$^\diamond$ & 0.0 & 290.80  \textrm{(fixed)} & 99.0(1) & 4.0  \textrm{(fixed)} \\
         & GGA+U$^\diamond$ & 0.0 & 299.00 & 90.9 & 4.3\\
         & GGA+U$^\star$ & 0.0 & 299.54 & 94.7 & - \\
         & EXP$^\dagger$ & 0.0 & 292.38 & 91.5 & 4.0   \textrm{(fixed)}  \\
         & GGA+U$^\dagger$ & 0.0  & 292.38 & 92.9 & -  \\
         & LDA+U$^\dagger$ & 0.0 & 300.62 & 114.1 & -  \\
         & EXP$^\ddagger$ & 0.0 & 291.60 & 106.0(8) & - \\
 \colrule
 HP-$Pbnm$ & EXP$^\diamond$ & 73.0 & 203.00  \textrm{(fixed)} &  545.0(2) & 4.0  \textrm{(fixed)} \\
 HP-$Pbnm$ (LS) & GGA+U$^\diamond$ & 73.0 &  196.47 &  381.0 & 4.0  \\
   \end{tabular}
\end{ruledtabular}
\end{table*}
\begin{figure}
\includegraphics[width=0.5\textwidth]{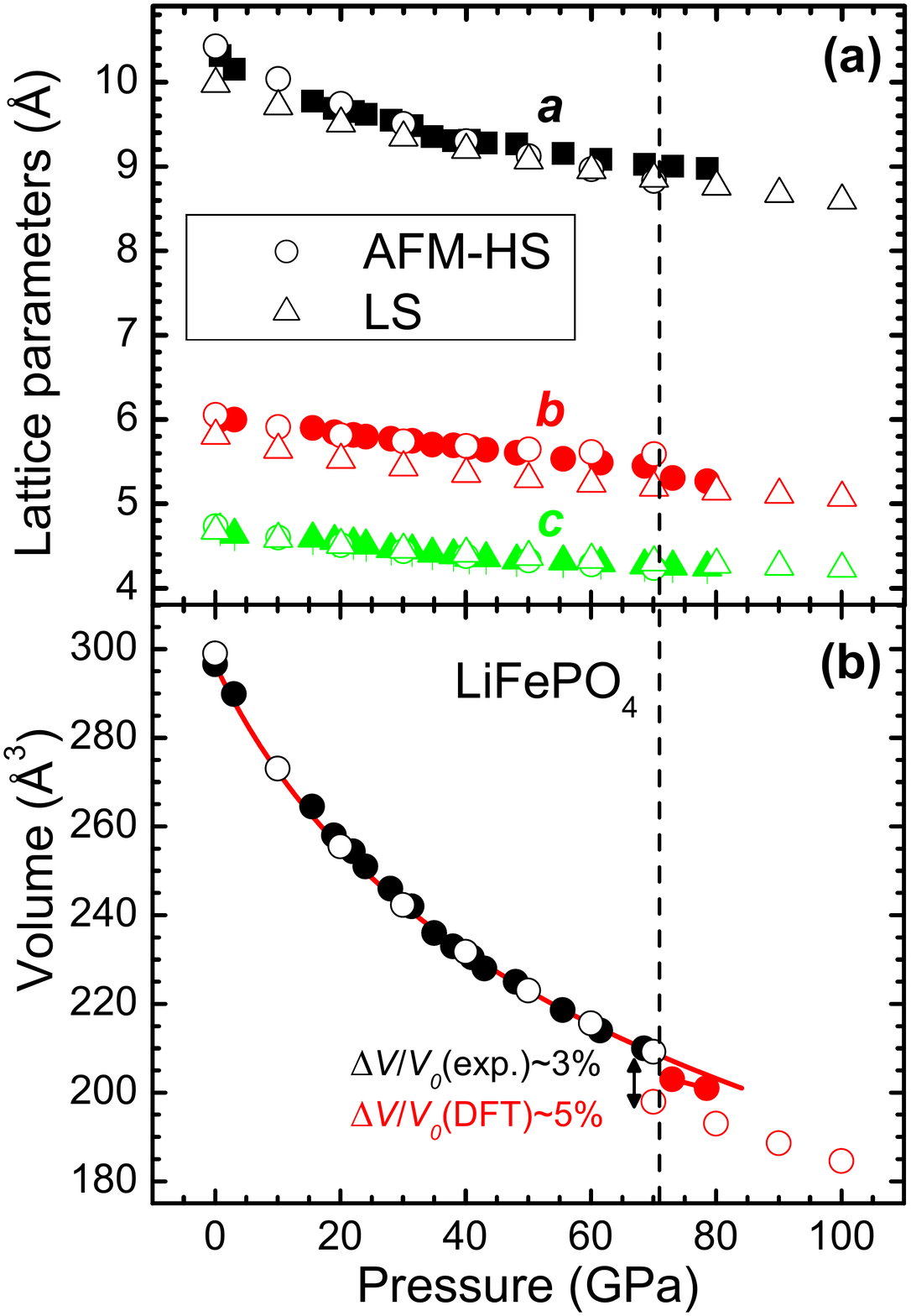}
\caption{\label{fig5} (Color online) (a) Lattice parameters (EXP: solid symbols, DFT+U: open symbols) and (b) unit cell-volume as a function of pressure for $Pbnm$--LiFePO$_{4}$ ($\lesssim73$ GPa; EXP: solid black circles, DFT+U: open black circles) and its HP-$Pbnm$ phase ($\gtrsim73$ GPa; EXP: solid red circles, DFT+U: open red circles). Solid lines running through the symbols are the fitted Birsch-Murnaghan EoS forms. The vertical dashed line marks the onset of the isostructural transition. Error bars lie within the symbols.}
\end{figure}
Following our vibrational spectroscopic results, we have additionally conducted high-pressure XRD experiments, in order to identify the aforementioned phase transition close to 70~GPa. We have only probed the structural behavior of sample B, since the Bragg peaks of the impurities in sample A (see Sec. II A, as well as the \textbf{Supplemental Material}) might hinder the clear detection of a structural change. Overall, the XRD patterns collected up to 79~GPa reveal indeed a transition initiating at 69~GPa, as exposed by the appearance of a shoulder in the Bragg peak centered at $2\theta\simeq7^{\circ}$ [Fig. 4(a)]. Further compression leads to the enhancement of this shoulder at the expense of the $7^{\circ}$ Bragg peak attributed to the starting $Pbnm$ phase.

Indexing of the XRD pattern at 78.5~GPa is possible again with a $Pbnm$ structure [Fig. 4(b)]. Hence, LiFePO$_{4}$ undergoes an isostructural transition close to 70~GPa, in excellent agreement with our vibrational studies discussed above. According to our extracted lattice parameters, Fig. 5(a), upon passing from the $Pbnm$ to the HP-$Pbnm$ phase, the orthorhombic $\mathbf{b}$-axis decreases by $\sim$3\%, which in turn results in an overall volume reduction of $\sim$3\% (Fig. 5). Thus, the isostructural $Pbnm$~$\rightarrow$~HP-$Pbnm$ transition can be classified as of first-order. We should also note that we tested the proposed $Cmcm$ phase\cite{Lin2011}, but it could not reproduce the observed high-pressure XRD patterns.

In Fig. 5 we plot the measured pressure-volume ($P-V$) data, alongside the respective equation of state (EoS) fittings and our GGA+U (with $U=2.5$ eV) results. The extracted elastic parameters are listed in Table~1. As we can observe, our volume at zero pressure, $\textit{V}_{0}$, and bulk modulus, $\textit{B}_{0}$, values for the starting $Pbnm$ phase of LiFePO$_{4}$ are in good agreement with the reported literature\cite{MaxischCeder2006,Dong2017,Dodd2007}.
\subsection{Structural relaxations}
\begin{figure}[h]
\includegraphics[width=0.4\textwidth]{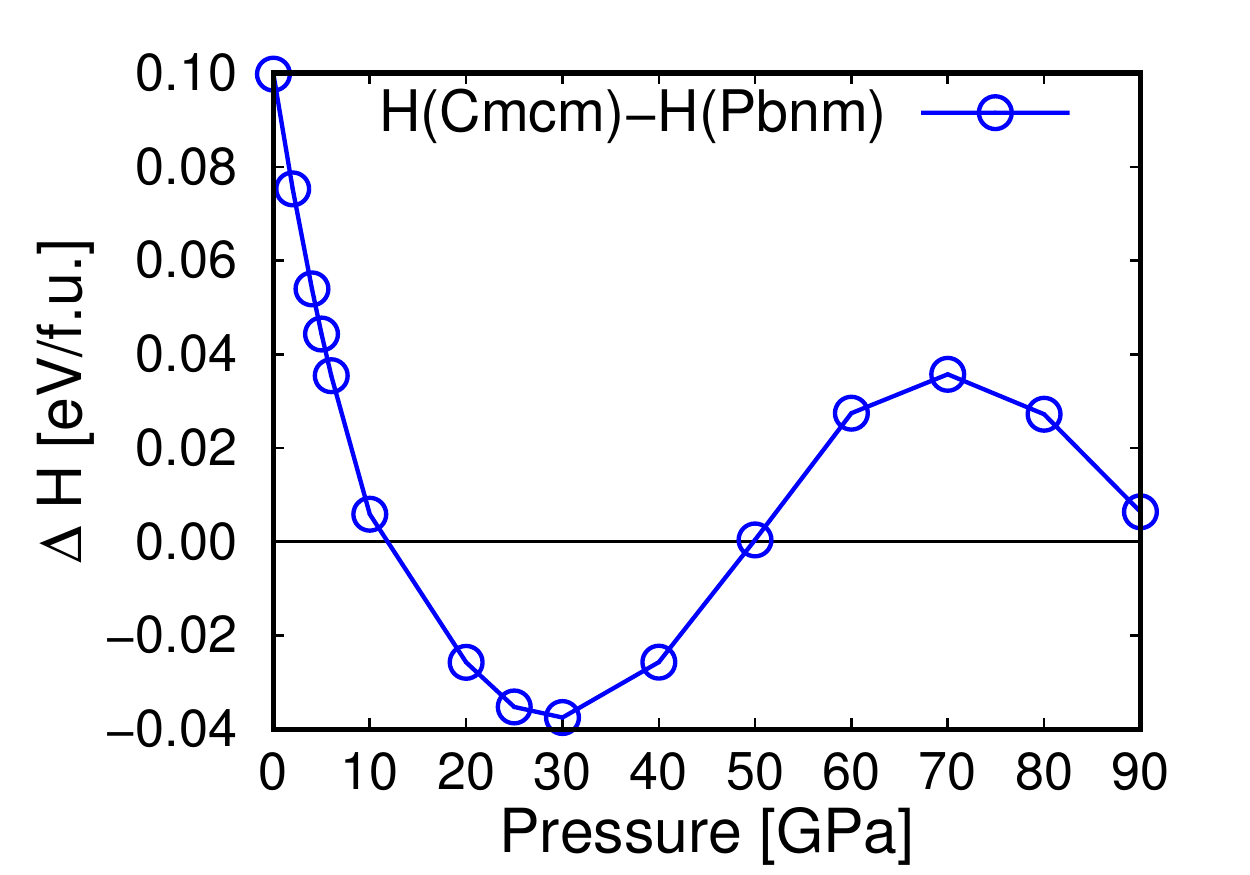}
\caption{\label{fig6} (Color online) Relative enthalpies of LiFePO$_4$ for the two AFM structures considered, $Pbnm$ and $Cmcm$.}
\end{figure}
In order to complement and verify our experimental results, we also performed DFT calculations. For this purpose, first we stabilized the AFM and FM configurations of the $Pbnm-$ and $Cmcm-$LiFePO$_4$ at ambient pressure, and determined the AFM ordering of $Pbnm-$LiFePO$_4$ to be the ground state at $\sim$7.5 meV and $\sim$12.5 meV per formula-unit (f.u.) lower than the FM $Pbnm-$LiFePO$_4$ and the AFM $Cmcm-$LiFePO$_4$ for $U=4.3$ eV, respectively. We began with $U=4.3$ eV as this value was used in the most recent work of stability of triphylite\cite{Dong2017}, and though $U$ seems to be not considered by Lin and Zeng\cite{Lin2011} in their $Pbnm\rightarrow Cmcm$ phase-transition study, the authors also report AFM ordering as the ground states for their $Pbnm-$ and $Cmcm-$triphylite phases. We note that AFM $Pbnm-$LiFePO$_4$ holds as the ground state independently of $U$ (=0, 2.5, 3.3, 4.3, 5.3 eV). The energy differences we found are consistent with first-principles calculations also performed in LiFePO$_4$ as part of a study of Li-compounds\cite{Zhou2004} at ambient conditions.  Thus, in the rest of the work we only examine AFM phases with $U=2.5$ eV, as it was the value to produce the most compatible structural parameters and description of the high-pressure spin state of LiFePO$_4$ (Fig. 5).

 To assess the stability of this orthorhombic  $Pbnm-$LiFePO$_4$ phase with respect to $Cmcm-$LiFePO$_4$, we calculated their enthalpy ($H$) as a function of pressure ($P$), volume ($V$), and internal energy (E) at each $P$,
\begin{equation}
  H(P)=E[V(P)]+PV(P).
\end{equation}
We found, that as pressure increases, the difference in enthalpy per f.u. ($\Delta H$) between the $Pbnm-$ and $Cmcm-$phases fluctuates for pressures between 0 and 90 GPa (Fig.~6), making these two competing phases almost degenerate within the accuracy of our static first-principles method. However, upon comparison with the structural XRD refinements [Fig.~4(a)], the $Pbnm$ structure is favored in this pressure range. Therefore, we conclude that the proposed pressure-induced $Pbnm\rightarrow Cmcm$ phase transition in LiFePO$_{4}$\cite{Lin2011} is not likely at RT. Further examination contrasting our experimental and calculated lattice parameters of LiFePO$_{4}$-triphylite  shows, in general, an excellent agreement for the starting AFM-HS $Pbnm$-phase ($\lesssim75$ GPa), Fig. 5(a) (see also Fig. 3 and Tables 3-5 in \textbf{Supplemental Material}). Upon reaching the transition pressure point, we notice that the magnitude of the lattice parameter along the $\mathbf{b}$-axis undergoes the largest decrease, thus contributing greatly to the volume drop (Fig. 5) observed experimentally ($\sim$3\%) and predicted by GGA+U ($\sim$5\%).  
\section{Spin Transition}
\begin{figure}
\includegraphics[width=0.5\textwidth]{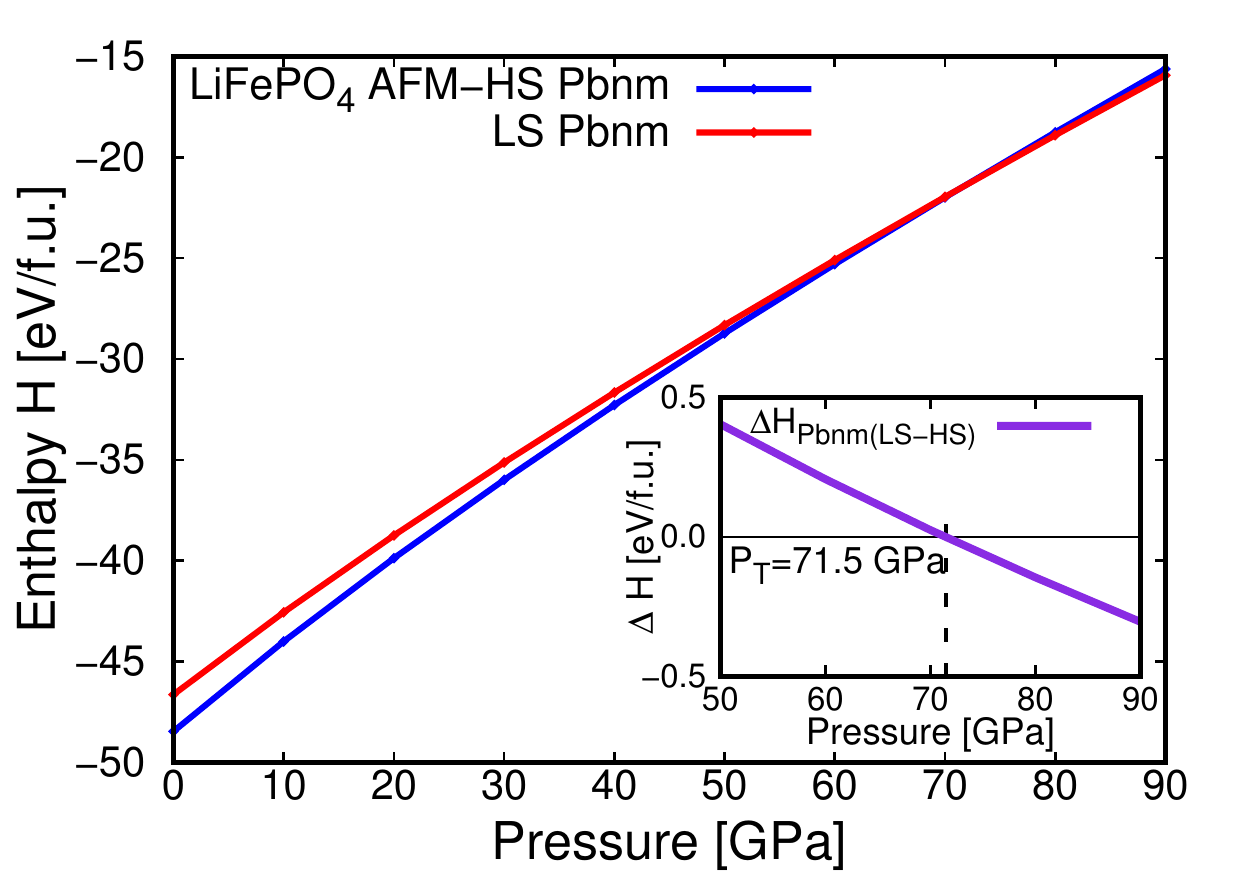}
\caption{\label{fig7} (Color online) Enthalphy of $Pbnm-$LiFePO$_4$ in HS and LS states as a function of pressure. The inset shows the spin transition pressure ($P_T$) as predicted from GGA+U, with $U=2.5$ eV.}
\end{figure}

\begin{figure}
\includegraphics[width=0.45\textwidth]{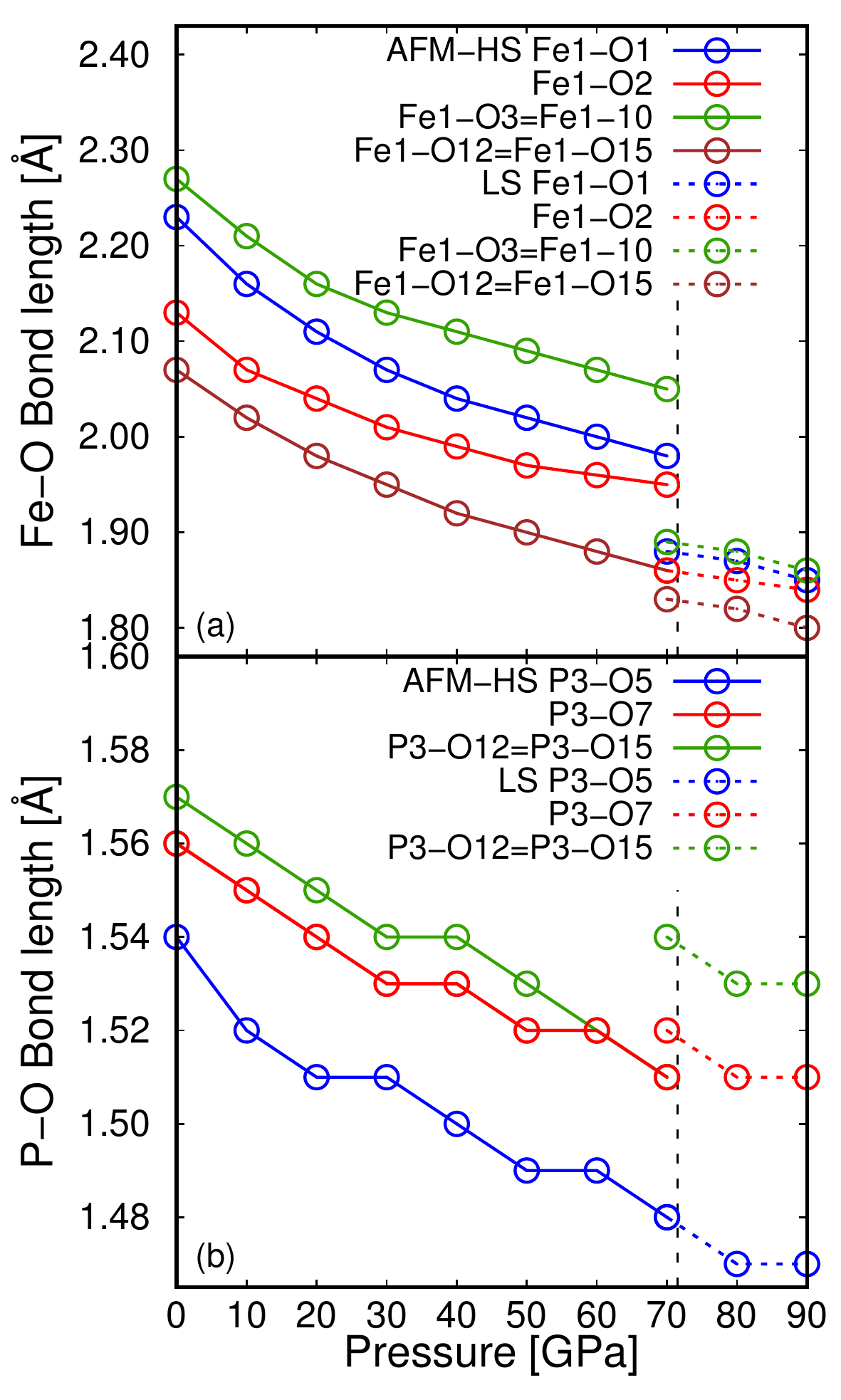}
\caption{\label{fig7} (Color online) Computed (a) Fe1-O and (b) P3-O bond length distances as a function of pressure for $Pbnm-$LiFePO$_4$ in HS and LS states within GGA+U, with $U=2.5$ eV. The vertical dashed lines mark the onset of the isostructural transition. Bond lengths for Fe2-O, Fe3-O, and Fe4-O and P1-O, P2-O, and P3-O are analogous (see Fig. 1). }
\end{figure}
As we have already mentioned in the Introduction (Sec. I), high-pressure studies of Fe-bearing minerals have established a HS-LS transition upon sufficient compression\cite{Badro2003,Badro2004,Mattila2007}. This magnetic change is usually accompanied by a first-order isostructural transition, i.e., the crystalline symmetry is retained during the HS-LS transition, whereas the first-order character is attributed to the reduction of the Fe$^{2+}$ ionic radius upon the spin change\cite{Shannon1969}. Using DFT we studied the spin state behavior of $Pbnm-$LiFePO$_4$ as a function of pressure within GGA+U with $U=2.5$ eV. We stabilized the $Pbnm-$LiFePO$_4$ in its LS state and computed its enthalpy (Eq. 2) to compare it to that of $Pbnm-$LiFePO$_4$ in AFM-HS configuration for pressures between 0 and 90 GPa. Fig. 7 shows the switch in spin state from HS-to-LS at about 72 GPa, in excellent agreement with the observed transition pressure in our experiments. To further corroborate the presence of this spin crossover, we also investigated the relative enthalpies of the HS and LS states using other $U$ values within the GGA+U and LDA+U schemes. Our results detailed in our \textbf{Supplemental Material} indicate that, the spin transition occurs for all values of $U$ tested (Fig. 4 in \textbf{Supplemental Material}), and that the spin transition pressure predicted by GGA+U increases with $U$. LDA+U also showed this spin-crossover, however, the structural parameters and volume of triphylite computed within this scheme were grossly underestimated (see Fig. 5 in \textbf{Supplemental Material}).

In order to gain a microscopic structural understanding of the HS-to-LS transition in LiFePO$_4$, we plot selected \textit{ab initio} Fe-O and P-O bond lengths as a function of pressure (Fig. 8). As we can clearly observe in Fig. 8(a), all the Fe-O bond distances shrink abrubtly at the transition point, following the decrease in the Fe$^{2+}$ ionic radius upon the spin crossover. This effect has been previously observed in other Fe-bearing compounds\cite{Lavina2009,Lavina2010,Lin2013}. At the same time, we notice that three of the P-O bonds increase across the transition pressure, Fig. 8(b). This P-O bond enlargement lies in excellent agreement with our Raman investigations, where a frequency downshift of the $\nu_1$ mode was observed at the transition point (Fig. 3). Since the $\nu_1$ mode corresponds to the symmetric P-O stretching motion\cite{Paques-LedentTarte1974,BurbaFrech2004}, its frequency downshift implies an increase of the associated P-O bonds at the spin crossover. Our DFT+U calculations have indeed captured this effect. 

Considering the very good consistency between our experimental and computational results, we can establish at this point that LiFePO$_{4}$ undergoes a \textit{spin-driven} isostructural transition at $\sim$70 GPa, with its magnetic configuration changing from AFM to non-magnetic owing to the pressure-induced HS-LS transition. Compared to relevant Fe-bearing minerals\cite{Badro2003,Lin2005,Speziale2005,Tsuchiya2006,Goncharov2006,Lin2007,Crowhurst2008,Wentzcovitch2009,Marquardt2009,Antonangeli2011,Wu2013,Holmstrom2015,Wu2014,Hsu2010,Badro2004,Li2004,McCammon2008,Lin2008,Hsu2010-pv,Hsu2011,Potapkin2013,Hsu2014,Mattila2007,Lavina2009,Lavina2010,Nagai2010,Farfan2012,Lin2012,Liu2014,Lobanov2015,Hsu2016,Muller2017}, LiFePO$_{4}$ exhibits the highest onset pressure for a spin transition to date. A plausible explanation might be the incompressibility of the FeO$_{6}$ octahedra in the olivine structure compared to, e.g., the ferropericlase and carbonate structures. In Fig. 8 we can see that as pressure increases, the Fe-O bond-length distance first decreases rather rapidily but at about 40 GPa, the contraction rate decreases.
\section{Conclusions}
We have investigated the olivine-type LiFePO$_4$ structural and magnetic properties under pressure, with a combination of first-principles and experimental probes. Our investigations indicate that the starting $Pbnm$ phase of LiFePO$_4$ persists up to $\sim$70~GPa. Further compression leads to a first-order isostructural transition in the pressure range of $\sim$70-75 GPa, inconsistent with previous claims regarding the adoption of a $Cmcm$ structure at $\sim$4 GPa\cite{Lin2011}. Considering our DFT results placing a high-spin to low-spin transition of Fe$^{2+}$ close to 72 GPa, we attribute the experimentally observed isostructural transition to a change of the spin state of Fe$^{2+}$ in LiFePO$_4$, i.e., a transition from the starting AFM-$Pbnm$ phase towards a non-magnetic state. To date, our results imply that LiFePO$_4$ exhibits the highest onset pressure for a spin state transition among Fe-bearing minerals, resulting most likely from the larger incompressibility of its FeO$_{6}$ octahedra.
\begin{acknowledgments}
We thank Dr. Catherine McCammon for the M\"{o}ssbauer measurements. This work was supported by the Deutsche Forschungs Gemeinschaft (DFG) funds Ko1260/18 and Wi2000/10. Parts of this research were carried out at the light source PETRA III at DESY, a member of the Helmholtz Association (HGF). MNV gratefully acknowledges the computing time granted by the John von Neumann Institute for Computing (NIC) and provided on the supercomputer JURECA at J\"ulich Supercomputing Centre (JSC) under project ID hpo24. Some computations were also performed at the GFZ linux cluster GLIC.
\end{acknowledgments}

\bibliography{MariNV_library}

\begin{thebibliography}{70}%
\makeatletter
\providecommand \@ifxundefined [1]{%
 \@ifx{#1\undefined}
}%
\providecommand \@ifnum [1]{%
 \ifnum #1\expandafter \@firstoftwo
 \else \expandafter \@secondoftwo
 \fi
}%
\providecommand \@ifx [1]{%
 \ifx #1\expandafter \@firstoftwo
 \else \expandafter \@secondoftwo
 \fi
}%
\providecommand \natexlab [1]{#1}%
\providecommand \enquote  [1]{``#1''}%
\providecommand \bibnamefont  [1]{#1}%
\providecommand \bibfnamefont [1]{#1}%
\providecommand \citenamefont [1]{#1}%
\providecommand \href@noop [0]{\@secondoftwo}%
\providecommand \href [0]{\begingroup \@sanitize@url \@href}%
\providecommand \@href[1]{\@@startlink{#1}\@@href}%
\providecommand \@@href[1]{\endgroup#1\@@endlink}%
\providecommand \@sanitize@url [0]{\catcode `\\12\catcode `\$12\catcode
  `\&12\catcode `\#12\catcode `\^12\catcode `\_12\catcode `\%12\relax}%
\providecommand \@@startlink[1]{}%
\providecommand \@@endlink[0]{}%
\providecommand \url  [0]{\begingroup\@sanitize@url \@url }%
\providecommand \@url [1]{\endgroup\@href {#1}{\urlprefix }}%
\providecommand \urlprefix  [0]{URL }%
\providecommand \Eprint [0]{\href }%
\providecommand \doibase [0]{http://dx.doi.org/}%
\providecommand \selectlanguage [0]{\@gobble}%
\providecommand \bibinfo  [0]{\@secondoftwo}%
\providecommand \bibfield  [0]{\@secondoftwo}%
\providecommand \translation [1]{[#1]}%
\providecommand \BibitemOpen [0]{}%
\providecommand \bibitemStop [0]{}%
\providecommand \bibitemNoStop [0]{.\EOS\space}%
\providecommand \EOS [0]{\spacefactor3000\relax}%
\providecommand \BibitemShut  [1]{\csname bibitem#1\endcsname}%
\let\auto@bib@innerbib\@empty
\bibitem [{\citenamefont {Losey}\ \emph {et~al.}(2004)\citenamefont {Losey},
  \citenamefont {Rakovan}, \citenamefont {Hughes}, \citenamefont {Francis},\
  and\ \citenamefont {Dyar}}]{Losey2004}%
  \BibitemOpen
  \bibfield  {author} {\bibinfo {author} {\bibfnamefont {A.}~\bibnamefont
  {Losey}}, \bibinfo {author} {\bibfnamefont {J.}~\bibnamefont {Rakovan}},
  \bibinfo {author} {\bibfnamefont {J.~M.}\ \bibnamefont {Hughes}}, \bibinfo
  {author} {\bibfnamefont {C.~A.}\ \bibnamefont {Francis}}, \ and\ \bibinfo
  {author} {\bibfnamefont {M.~D.}\ \bibnamefont {Dyar}},\ }\href@noop {}
  {\bibfield  {journal} {\bibinfo  {journal} {Can. Mineral.}\ }\textbf
  {\bibinfo {volume} {42}},\ \bibinfo {pages} {1105} (\bibinfo {year}
  {2004})}\BibitemShut {NoStop}%
\bibitem [{\citenamefont {Ringwood}(1975)}]{Ringwood1975}%
  \BibitemOpen
  \bibfield  {author} {\bibinfo {author} {\bibfnamefont {A.~E.}\ \bibnamefont
  {Ringwood}},\ }\href@noop {} {\emph {\bibinfo {title} {Composition and
  Petrology of the Earth's Mantle}}}\ (\bibinfo  {publisher} {McGraw‐Hill},\
  \bibinfo {address} {New York, NY, USA},\ \bibinfo {year} {1975})\ p.\
  \bibinfo {pages} {618}\BibitemShut {NoStop}%
\bibitem [{\citenamefont {Badro}(2003)}]{Badro2003}%
  \BibitemOpen
  \bibfield  {author} {\bibinfo {author} {\bibfnamefont {J.}~\bibnamefont
  {Badro}},\ }\href {\doibase 10.1126/science.1081311} {\bibfield  {journal}
  {\bibinfo  {journal} {Science}\ }\textbf {\bibinfo {volume} {300}},\ \bibinfo
  {pages} {789} (\bibinfo {year} {2003})}\BibitemShut {NoStop}%
\bibitem [{\citenamefont {Lin}\ \emph {et~al.}(2005)\citenamefont {Lin},
  \citenamefont {Struzhkin}, \citenamefont {Jacobsen}, \citenamefont {Hu},
  \citenamefont {Chow}, \citenamefont {Kung}, \citenamefont {Liu},
  \citenamefont {Mao},\ and\ \citenamefont {Hemley}}]{Lin2005}%
  \BibitemOpen
  \bibfield  {author} {\bibinfo {author} {\bibfnamefont {J.-F.}\ \bibnamefont
  {Lin}}, \bibinfo {author} {\bibfnamefont {V.~V.}\ \bibnamefont {Struzhkin}},
  \bibinfo {author} {\bibfnamefont {S.~D.}\ \bibnamefont {Jacobsen}}, \bibinfo
  {author} {\bibfnamefont {M.~Y.}\ \bibnamefont {Hu}}, \bibinfo {author}
  {\bibfnamefont {P.}~\bibnamefont {Chow}}, \bibinfo {author} {\bibfnamefont
  {J.}~\bibnamefont {Kung}}, \bibinfo {author} {\bibfnamefont {H.}~\bibnamefont
  {Liu}}, \bibinfo {author} {\bibfnamefont {H.-k.}\ \bibnamefont {Mao}}, \ and\
  \bibinfo {author} {\bibfnamefont {R.~J.}\ \bibnamefont {Hemley}},\ }\href
  {\doibase 10.1038/nature03825} {\bibfield  {journal} {\bibinfo  {journal}
  {Nature}\ }\textbf {\bibinfo {volume} {436}},\ \bibinfo {pages} {377}
  (\bibinfo {year} {2005})}\BibitemShut {NoStop}%
\bibitem [{\citenamefont {Speziale}\ \emph {et~al.}(2005)\citenamefont
  {Speziale}, \citenamefont {Milner}, \citenamefont {Lee}, \citenamefont
  {Clark}, \citenamefont {Pasternak},\ and\ \citenamefont
  {Jeanloz}}]{Speziale2005}%
  \BibitemOpen
  \bibfield  {author} {\bibinfo {author} {\bibfnamefont {S.}~\bibnamefont
  {Speziale}}, \bibinfo {author} {\bibfnamefont {a.}~\bibnamefont {Milner}},
  \bibinfo {author} {\bibfnamefont {V.~E.}\ \bibnamefont {Lee}}, \bibinfo
  {author} {\bibfnamefont {S.~M.}\ \bibnamefont {Clark}}, \bibinfo {author}
  {\bibfnamefont {M.~P.}\ \bibnamefont {Pasternak}}, \ and\ \bibinfo {author}
  {\bibfnamefont {R.}~\bibnamefont {Jeanloz}},\ }\href {\doibase
  10.1073/pnas.0508919102} {\bibfield  {journal} {\bibinfo  {journal} {Proc.
  Natl. Acad. Sci. USA}\ }\textbf {\bibinfo {volume} {102}},\ \bibinfo {pages}
  {17918} (\bibinfo {year} {2005})}\BibitemShut {NoStop}%
\bibitem [{\citenamefont {Tsuchiya}\ \emph {et~al.}(2006)\citenamefont
  {Tsuchiya}, \citenamefont {Wentzcovitch}, \citenamefont {Da~Silva},\ and\
  \citenamefont {{de Gironcoli}}}]{Tsuchiya2006}%
  \BibitemOpen
  \bibfield  {author} {\bibinfo {author} {\bibfnamefont {T.}~\bibnamefont
  {Tsuchiya}}, \bibinfo {author} {\bibfnamefont {R.~M.}\ \bibnamefont
  {Wentzcovitch}}, \bibinfo {author} {\bibfnamefont {C.~R.~S.}\ \bibnamefont
  {Da~Silva}}, \ and\ \bibinfo {author} {\bibfnamefont {S.}~\bibnamefont {{de
  Gironcoli}}},\ }\href {\doibase 10.1103/PhysRevLett.96.198501} {\bibfield
  {journal} {\bibinfo  {journal} {Phys. Rev. Lett.}\ }\textbf {\bibinfo
  {volume} {96}},\ \bibinfo {pages} {1} (\bibinfo {year} {2006})}\BibitemShut
  {NoStop}%
\bibitem [{\citenamefont {Goncharov}\ \emph {et~al.}(2006)\citenamefont
  {Goncharov}, \citenamefont {Struzhkin},\ and\ \citenamefont
  {Jacobsen}}]{Goncharov2006}%
  \BibitemOpen
  \bibfield  {author} {\bibinfo {author} {\bibfnamefont {A.~F.}\ \bibnamefont
  {Goncharov}}, \bibinfo {author} {\bibfnamefont {V.~V.}\ \bibnamefont
  {Struzhkin}}, \ and\ \bibinfo {author} {\bibfnamefont {S.~D.}\ \bibnamefont
  {Jacobsen}},\ }\href {\doibase 10.1126/science.1125622} {\bibfield  {journal}
  {\bibinfo  {journal} {Science}\ }\textbf {\bibinfo {volume} {312}},\ \bibinfo
  {pages} {1205} (\bibinfo {year} {2006})}\BibitemShut {NoStop}%
\bibitem [{\citenamefont {Lin}\ \emph {et~al.}(2007)\citenamefont {Lin},
  \citenamefont {Vanko}, \citenamefont {Jacobsen}, \citenamefont {Iota},
  \citenamefont {Struzhkin}, \citenamefont {Prakapenka}, \citenamefont
  {Kuznetsov},\ and\ \citenamefont {Yoo}}]{Lin2007}%
  \BibitemOpen
  \bibfield  {author} {\bibinfo {author} {\bibfnamefont {J.-F.}\ \bibnamefont
  {Lin}}, \bibinfo {author} {\bibfnamefont {G.}~\bibnamefont {Vanko}}, \bibinfo
  {author} {\bibfnamefont {S.~D.}\ \bibnamefont {Jacobsen}}, \bibinfo {author}
  {\bibfnamefont {V.}~\bibnamefont {Iota}}, \bibinfo {author} {\bibfnamefont
  {V.~V.}\ \bibnamefont {Struzhkin}}, \bibinfo {author} {\bibfnamefont {V.~B.}\
  \bibnamefont {Prakapenka}}, \bibinfo {author} {\bibfnamefont
  {A.}~\bibnamefont {Kuznetsov}}, \ and\ \bibinfo {author} {\bibfnamefont
  {C.-S.}\ \bibnamefont {Yoo}},\ }\href {\doibase 10.1126/science.1144997}
  {\bibfield  {journal} {\bibinfo  {journal} {Science}\ }\textbf {\bibinfo
  {volume} {317}},\ \bibinfo {pages} {1740} (\bibinfo {year}
  {2007})}\BibitemShut {NoStop}%
\bibitem [{\citenamefont {Crowhurst}\ \emph {et~al.}(2008)\citenamefont
  {Crowhurst}, \citenamefont {Brown}, \citenamefont {Goncharov},\ and\
  \citenamefont {Jacobsen}}]{Crowhurst2008}%
  \BibitemOpen
  \bibfield  {author} {\bibinfo {author} {\bibfnamefont {J.~C.}\ \bibnamefont
  {Crowhurst}}, \bibinfo {author} {\bibfnamefont {J.~M.}\ \bibnamefont
  {Brown}}, \bibinfo {author} {\bibfnamefont {a.~F.}\ \bibnamefont
  {Goncharov}}, \ and\ \bibinfo {author} {\bibfnamefont {S.~D.}\ \bibnamefont
  {Jacobsen}},\ }\href {\doibase 10.1126/science.1149606} {\bibfield  {journal}
  {\bibinfo  {journal} {Science}\ }\textbf {\bibinfo {volume} {319}},\ \bibinfo
  {pages} {451} (\bibinfo {year} {2008})}\BibitemShut {NoStop}%
\bibitem [{\citenamefont {Wentzcovitch}\ \emph {et~al.}(2009)\citenamefont
  {Wentzcovitch}, \citenamefont {Justo}, \citenamefont {Wu}, \citenamefont
  {da~Silva}, \citenamefont {Yuen},\ and\ \citenamefont
  {Kohlstedt}}]{Wentzcovitch2009}%
  \BibitemOpen
  \bibfield  {author} {\bibinfo {author} {\bibfnamefont {R.~M.}\ \bibnamefont
  {Wentzcovitch}}, \bibinfo {author} {\bibfnamefont {J.~F.}\ \bibnamefont
  {Justo}}, \bibinfo {author} {\bibfnamefont {Z.}~\bibnamefont {Wu}}, \bibinfo
  {author} {\bibfnamefont {C.~R.~S.}\ \bibnamefont {da~Silva}}, \bibinfo
  {author} {\bibfnamefont {D.~A.}\ \bibnamefont {Yuen}}, \ and\ \bibinfo
  {author} {\bibfnamefont {D.}~\bibnamefont {Kohlstedt}},\ }\href {\doibase
  10.1073/pnas.0812150106} {\bibfield  {journal} {\bibinfo  {journal} {Proc.
  Natl. Acad. Sci. USA}\ }\textbf {\bibinfo {volume} {106}},\ \bibinfo {pages}
  {8447} (\bibinfo {year} {2009})}\BibitemShut {NoStop}%
\bibitem [{\citenamefont {Marquardt}\ \emph {et~al.}(2009)\citenamefont
  {Marquardt}, \citenamefont {Speziale}, \citenamefont {Reichmann},
  \citenamefont {Frost}, \citenamefont {Schilling},\ and\ \citenamefont
  {Garnero}}]{Marquardt2009}%
  \BibitemOpen
  \bibfield  {author} {\bibinfo {author} {\bibfnamefont {H.}~\bibnamefont
  {Marquardt}}, \bibinfo {author} {\bibfnamefont {S.}~\bibnamefont {Speziale}},
  \bibinfo {author} {\bibfnamefont {H.~J.}\ \bibnamefont {Reichmann}}, \bibinfo
  {author} {\bibfnamefont {D.~J.}\ \bibnamefont {Frost}}, \bibinfo {author}
  {\bibfnamefont {F.~R.}\ \bibnamefont {Schilling}}, \ and\ \bibinfo {author}
  {\bibfnamefont {E.~J.}\ \bibnamefont {Garnero}},\ }\href {\doibase
  10.1126/science.1169365} {\bibfield  {journal} {\bibinfo  {journal}
  {Science}\ }\textbf {\bibinfo {volume} {324}},\ \bibinfo {pages} {224}
  (\bibinfo {year} {2009})}\BibitemShut {NoStop}%
\bibitem [{\citenamefont {Antonangeli}\ \emph {et~al.}(2011)\citenamefont
  {Antonangeli}, \citenamefont {Siebert}, \citenamefont {Aracne}, \citenamefont
  {Farber}, \citenamefont {Bosak}, \citenamefont {Hoesch}, \citenamefont
  {Krisch}, \citenamefont {Ryerson}, \citenamefont {Fiquet},\ and\
  \citenamefont {Badro}}]{Antonangeli2011}%
  \BibitemOpen
  \bibfield  {author} {\bibinfo {author} {\bibfnamefont {D.}~\bibnamefont
  {Antonangeli}}, \bibinfo {author} {\bibfnamefont {J.}~\bibnamefont
  {Siebert}}, \bibinfo {author} {\bibfnamefont {C.~M.}\ \bibnamefont {Aracne}},
  \bibinfo {author} {\bibfnamefont {D.~L.}\ \bibnamefont {Farber}}, \bibinfo
  {author} {\bibfnamefont {A.}~\bibnamefont {Bosak}}, \bibinfo {author}
  {\bibfnamefont {M.}~\bibnamefont {Hoesch}}, \bibinfo {author} {\bibfnamefont
  {M.}~\bibnamefont {Krisch}}, \bibinfo {author} {\bibfnamefont {F.~J.}\
  \bibnamefont {Ryerson}}, \bibinfo {author} {\bibfnamefont {G.}~\bibnamefont
  {Fiquet}}, \ and\ \bibinfo {author} {\bibfnamefont {J.}~\bibnamefont
  {Badro}},\ }\href {\doibase 10.1126/science.1198429} {\bibfield  {journal}
  {\bibinfo  {journal} {Science}\ }\textbf {\bibinfo {volume} {331}},\ \bibinfo
  {pages} {64} (\bibinfo {year} {2011})}\BibitemShut {NoStop}%
\bibitem [{\citenamefont {Wu}\ \emph {et~al.}(2013)\citenamefont {Wu},
  \citenamefont {Justo},\ and\ \citenamefont {Wentzcovitch}}]{Wu2013}%
  \BibitemOpen
  \bibfield  {author} {\bibinfo {author} {\bibfnamefont {Z.}~\bibnamefont
  {Wu}}, \bibinfo {author} {\bibfnamefont {J.~F.}\ \bibnamefont {Justo}}, \
  and\ \bibinfo {author} {\bibfnamefont {R.~M.}\ \bibnamefont {Wentzcovitch}},\
  }\href {\doibase 10.1103/PhysRevLett.110.228501} {\bibfield  {journal}
  {\bibinfo  {journal} {Phys. Rev. Lett.}\ }\textbf {\bibinfo {volume} {110}},\
  \bibinfo {pages} {1} (\bibinfo {year} {2013})}\BibitemShut {NoStop}%
\bibitem [{\citenamefont {Holmstr{\"{o}}m}\ and\ \citenamefont
  {Stixrude}(2015)}]{Holmstrom2015}%
  \BibitemOpen
  \bibfield  {author} {\bibinfo {author} {\bibfnamefont {E.}~\bibnamefont
  {Holmstr{\"{o}}m}}\ and\ \bibinfo {author} {\bibfnamefont {L.}~\bibnamefont
  {Stixrude}},\ }\href {\doibase 10.1103/PhysRevLett.114.117202} {\bibfield
  {journal} {\bibinfo  {journal} {Phys. Rev. Lett.}\ }\textbf {\bibinfo
  {volume} {114}},\ \bibinfo {pages} {1} (\bibinfo {year} {2015})}\BibitemShut
  {NoStop}%
\bibitem [{\citenamefont {Wu}\ and\ \citenamefont
  {Wentzcovitch}(2014)}]{Wu2014}%
  \BibitemOpen
  \bibfield  {author} {\bibinfo {author} {\bibfnamefont {Z.}~\bibnamefont
  {Wu}}\ and\ \bibinfo {author} {\bibfnamefont {R.~M.}\ \bibnamefont
  {Wentzcovitch}},\ }\href {\doibase 10.1073/pnas.1322427111} {\bibfield
  {journal} {\bibinfo  {journal} {Proc. Natl. Acad. Sci. USA}\ }\textbf
  {\bibinfo {volume} {111}},\ \bibinfo {pages} {10468} (\bibinfo {year}
  {2014})}\BibitemShut {NoStop}%
\bibitem [{\citenamefont {Hsu}\ \emph {et~al.}(2010{\natexlab{a}})\citenamefont
  {Hsu}, \citenamefont {Umemoto}, \citenamefont {Wu},\ and\ \citenamefont
  {Wentzcovitch}}]{Hsu2010}%
  \BibitemOpen
  \bibfield  {author} {\bibinfo {author} {\bibfnamefont {H.}~\bibnamefont
  {Hsu}}, \bibinfo {author} {\bibfnamefont {K.}~\bibnamefont {Umemoto}},
  \bibinfo {author} {\bibfnamefont {Z.}~\bibnamefont {Wu}}, \ and\ \bibinfo
  {author} {\bibfnamefont {R.~M.}\ \bibnamefont {Wentzcovitch}},\ }\href
  {\doibase 10.2138/rmg.2010.71.09} {\bibfield  {journal} {\bibinfo  {journal}
  {Rev. Mineral. Geochemistry}\ }\textbf {\bibinfo {volume} {71}},\ \bibinfo
  {pages} {169} (\bibinfo {year} {2010}{\natexlab{a}})}\BibitemShut {NoStop}%
\bibitem [{\citenamefont {Badro}(2004)}]{Badro2004}%
  \BibitemOpen
  \bibfield  {author} {\bibinfo {author} {\bibfnamefont {J.}~\bibnamefont
  {Badro}},\ }\href {\doibase 10.1126/science.1098840} {\bibfield  {journal}
  {\bibinfo  {journal} {Science}\ }\textbf {\bibinfo {volume} {305}},\ \bibinfo
  {pages} {383} (\bibinfo {year} {2004})}\BibitemShut {NoStop}%
\bibitem [{\citenamefont {Li}\ \emph {et~al.}(2004)\citenamefont {Li},
  \citenamefont {Struzhkin}, \citenamefont {Mao}, \citenamefont {Shu},
  \citenamefont {Hemley}, \citenamefont {Fei}, \citenamefont {Mysen},
  \citenamefont {Dera}, \citenamefont {Prakapenka},\ and\ \citenamefont
  {Shen}}]{Li2004}%
  \BibitemOpen
  \bibfield  {author} {\bibinfo {author} {\bibfnamefont {J.}~\bibnamefont
  {Li}}, \bibinfo {author} {\bibfnamefont {V.~V.}\ \bibnamefont {Struzhkin}},
  \bibinfo {author} {\bibfnamefont {H.-K.}\ \bibnamefont {Mao}}, \bibinfo
  {author} {\bibfnamefont {J.}~\bibnamefont {Shu}}, \bibinfo {author}
  {\bibfnamefont {R.~J.}\ \bibnamefont {Hemley}}, \bibinfo {author}
  {\bibfnamefont {Y.}~\bibnamefont {Fei}}, \bibinfo {author} {\bibfnamefont
  {B.}~\bibnamefont {Mysen}}, \bibinfo {author} {\bibfnamefont
  {P.}~\bibnamefont {Dera}}, \bibinfo {author} {\bibfnamefont {V.}~\bibnamefont
  {Prakapenka}}, \ and\ \bibinfo {author} {\bibfnamefont {G.}~\bibnamefont
  {Shen}},\ }\href {\doibase 10.1073/pnas.0405804101} {\bibfield  {journal}
  {\bibinfo  {journal} {Proc. Natl. Acad. Sci. U. S. A.}\ }\textbf {\bibinfo
  {volume} {101}},\ \bibinfo {pages} {14027} (\bibinfo {year}
  {2004})}\BibitemShut {NoStop}%
\bibitem [{\citenamefont {McCammon}\ \emph {et~al.}(2008)\citenamefont
  {McCammon}, \citenamefont {Kantor}, \citenamefont {Narygina}, \citenamefont
  {Rouquette}, \citenamefont {Ponkratz}, \citenamefont {Sergueev},
  \citenamefont {Mezouar}, \citenamefont {Prakapenka},\ and\ \citenamefont
  {Dubrovinsky}}]{McCammon2008}%
  \BibitemOpen
  \bibfield  {author} {\bibinfo {author} {\bibfnamefont {C.}~\bibnamefont
  {McCammon}}, \bibinfo {author} {\bibfnamefont {I.}~\bibnamefont {Kantor}},
  \bibinfo {author} {\bibfnamefont {O.}~\bibnamefont {Narygina}}, \bibinfo
  {author} {\bibfnamefont {J.}~\bibnamefont {Rouquette}}, \bibinfo {author}
  {\bibfnamefont {U.}~\bibnamefont {Ponkratz}}, \bibinfo {author}
  {\bibfnamefont {I.}~\bibnamefont {Sergueev}}, \bibinfo {author}
  {\bibfnamefont {M.}~\bibnamefont {Mezouar}}, \bibinfo {author} {\bibfnamefont
  {V.}~\bibnamefont {Prakapenka}}, \ and\ \bibinfo {author} {\bibfnamefont
  {L.}~\bibnamefont {Dubrovinsky}},\ }\href {\doibase 10.1038/ngeo309}
  {\bibfield  {journal} {\bibinfo  {journal} {Nat. Geosci.}\ }\textbf {\bibinfo
  {volume} {1}},\ \bibinfo {pages} {684} (\bibinfo {year} {2008})}\BibitemShut
  {NoStop}%
\bibitem [{\citenamefont {Lin}\ \emph {et~al.}(2008)\citenamefont {Lin},
  \citenamefont {Watson}, \citenamefont {Vank{\'{o}}}, \citenamefont {Alp},
  \citenamefont {Prakapenka}, \citenamefont {Dera}, \citenamefont {Struzhkin},
  \citenamefont {Kubo}, \citenamefont {Zhao}, \citenamefont {McCammon},\ and\
  \citenamefont {Evans}}]{Lin2008}%
  \BibitemOpen
  \bibfield  {author} {\bibinfo {author} {\bibfnamefont {J.-F.}\ \bibnamefont
  {Lin}}, \bibinfo {author} {\bibfnamefont {H.}~\bibnamefont {Watson}},
  \bibinfo {author} {\bibfnamefont {G.}~\bibnamefont {Vank{\'{o}}}}, \bibinfo
  {author} {\bibfnamefont {E.~E.}\ \bibnamefont {Alp}}, \bibinfo {author}
  {\bibfnamefont {V.~B.}\ \bibnamefont {Prakapenka}}, \bibinfo {author}
  {\bibfnamefont {P.}~\bibnamefont {Dera}}, \bibinfo {author} {\bibfnamefont
  {V.~V.}\ \bibnamefont {Struzhkin}}, \bibinfo {author} {\bibfnamefont
  {A.}~\bibnamefont {Kubo}}, \bibinfo {author} {\bibfnamefont {J.}~\bibnamefont
  {Zhao}}, \bibinfo {author} {\bibfnamefont {C.}~\bibnamefont {McCammon}}, \
  and\ \bibinfo {author} {\bibfnamefont {W.~J.}\ \bibnamefont {Evans}},\ }\href
  {\doibase 10.1038/ngeo310} {\bibfield  {journal} {\bibinfo  {journal} {Nat.
  Geosci.}\ }\textbf {\bibinfo {volume} {1}},\ \bibinfo {pages} {688} (\bibinfo
  {year} {2008})}\BibitemShut {NoStop}%
\bibitem [{\citenamefont {Hsu}\ \emph {et~al.}(2010{\natexlab{b}})\citenamefont
  {Hsu}, \citenamefont {Umemoto}, \citenamefont {Blaha},\ and\ \citenamefont
  {Wentzcovitch}}]{Hsu2010-pv}%
  \BibitemOpen
  \bibfield  {author} {\bibinfo {author} {\bibfnamefont {H.}~\bibnamefont
  {Hsu}}, \bibinfo {author} {\bibfnamefont {K.}~\bibnamefont {Umemoto}},
  \bibinfo {author} {\bibfnamefont {P.}~\bibnamefont {Blaha}}, \ and\ \bibinfo
  {author} {\bibfnamefont {R.~M.}\ \bibnamefont {Wentzcovitch}},\ }\href
  {\doibase 10.1016/j.epsl.2010.02.031} {\bibfield  {journal} {\bibinfo
  {journal} {Earth Planet. Sci. Lett.}\ }\textbf {\bibinfo {volume} {294}},\
  \bibinfo {pages} {19} (\bibinfo {year} {2010}{\natexlab{b}})}\BibitemShut
  {NoStop}%
\bibitem [{\citenamefont {Hsu}\ \emph {et~al.}(2011)\citenamefont {Hsu},
  \citenamefont {Blaha}, \citenamefont {Cococcioni},\ and\ \citenamefont
  {Wentzcovitch}}]{Hsu2011}%
  \BibitemOpen
  \bibfield  {author} {\bibinfo {author} {\bibfnamefont {H.}~\bibnamefont
  {Hsu}}, \bibinfo {author} {\bibfnamefont {P.}~\bibnamefont {Blaha}}, \bibinfo
  {author} {\bibfnamefont {M.}~\bibnamefont {Cococcioni}}, \ and\ \bibinfo
  {author} {\bibfnamefont {R.~M.}\ \bibnamefont {Wentzcovitch}},\ }\href@noop
  {} {\bibfield  {journal} {\bibinfo  {journal} {Phys. Rev. Lett.}\ }\textbf
  {\bibinfo {volume} {106}},\ \bibinfo {pages} {118501} (\bibinfo {year}
  {2011})}\BibitemShut {NoStop}%
\bibitem [{\citenamefont {Potapkin}\ \emph {et~al.}(2013)\citenamefont
  {Potapkin}, \citenamefont {McCammon}, \citenamefont {Glazyrin}, \citenamefont
  {Kantor}, \citenamefont {Kupenko}, \citenamefont {Prescher}, \citenamefont
  {Sinmyo}, \citenamefont {Smirnov}, \citenamefont {Chumakov}, \citenamefont
  {R{\"{u}}ffer},\ and\ \citenamefont {Dubrovinsky}}]{Potapkin2013}%
  \BibitemOpen
  \bibfield  {author} {\bibinfo {author} {\bibfnamefont {V.}~\bibnamefont
  {Potapkin}}, \bibinfo {author} {\bibfnamefont {C.}~\bibnamefont {McCammon}},
  \bibinfo {author} {\bibfnamefont {K.}~\bibnamefont {Glazyrin}}, \bibinfo
  {author} {\bibfnamefont {A.}~\bibnamefont {Kantor}}, \bibinfo {author}
  {\bibfnamefont {I.}~\bibnamefont {Kupenko}}, \bibinfo {author} {\bibfnamefont
  {C.}~\bibnamefont {Prescher}}, \bibinfo {author} {\bibfnamefont
  {R.}~\bibnamefont {Sinmyo}}, \bibinfo {author} {\bibfnamefont {G.~V.}\
  \bibnamefont {Smirnov}}, \bibinfo {author} {\bibfnamefont {A.~I.}\
  \bibnamefont {Chumakov}}, \bibinfo {author} {\bibfnamefont {R.}~\bibnamefont
  {R{\"{u}}ffer}}, \ and\ \bibinfo {author} {\bibfnamefont {L.}~\bibnamefont
  {Dubrovinsky}},\ }\href {\doibase 10.1038/ncomms2436} {\bibfield  {journal}
  {\bibinfo  {journal} {Nat. Commun.}\ }\textbf {\bibinfo {volume} {4}},\
  \bibinfo {pages} {1427} (\bibinfo {year} {2013})}\BibitemShut {NoStop}%
\bibitem [{\citenamefont {Hsu}\ and\ \citenamefont
  {Wentzcovitch}(2014)}]{Hsu2014}%
  \BibitemOpen
  \bibfield  {author} {\bibinfo {author} {\bibfnamefont {H.}~\bibnamefont
  {Hsu}}\ and\ \bibinfo {author} {\bibfnamefont {R.~M.}\ \bibnamefont
  {Wentzcovitch}},\ }\href@noop {} {\bibfield  {journal} {\bibinfo  {journal}
  {Phys. Rev. B}\ }\textbf {\bibinfo {volume} {90}},\ \bibinfo {pages} {195205}
  (\bibinfo {year} {2014})}\BibitemShut {NoStop}%
\bibitem [{\citenamefont {Mattila}\ and\ \citenamefont
  {Pylkk{\"{a}}nen}(2007)}]{Mattila2007}%
  \BibitemOpen
  \bibfield  {author} {\bibinfo {author} {\bibfnamefont {A.}~\bibnamefont
  {Mattila}}\ and\ \bibinfo {author} {\bibfnamefont {T.}~\bibnamefont
  {Pylkk{\"{a}}nen}},\ }\href {\doibase 10.1088/0953-8984/19/38/386206}
  {\bibfield  {journal} {\bibinfo  {journal} {J. Phys. Condens. Matter}\
  }\textbf {\bibinfo {volume} {19}},\ \bibinfo {pages} {386206} (\bibinfo
  {year} {2007})}\BibitemShut {NoStop}%
\bibitem [{\citenamefont {Lavina}\ \emph {et~al.}(2009)\citenamefont {Lavina},
  \citenamefont {Dera}, \citenamefont {Downs}, \citenamefont {Prakapenka},
  \citenamefont {Rivers}, \citenamefont {Sutton},\ and\ \citenamefont
  {Nicol}}]{Lavina2009}%
  \BibitemOpen
  \bibfield  {author} {\bibinfo {author} {\bibfnamefont {B.}~\bibnamefont
  {Lavina}}, \bibinfo {author} {\bibfnamefont {P.}~\bibnamefont {Dera}},
  \bibinfo {author} {\bibfnamefont {R.~T.}\ \bibnamefont {Downs}}, \bibinfo
  {author} {\bibfnamefont {V.}~\bibnamefont {Prakapenka}}, \bibinfo {author}
  {\bibfnamefont {M.}~\bibnamefont {Rivers}}, \bibinfo {author} {\bibfnamefont
  {S.}~\bibnamefont {Sutton}}, \ and\ \bibinfo {author} {\bibfnamefont
  {M.}~\bibnamefont {Nicol}},\ }\href {\doibase 10.1029/2009GL039652}
  {\bibfield  {journal} {\bibinfo  {journal} {Geophys. Res. Lett.}\ }\textbf
  {\bibinfo {volume} {36}},\ \bibinfo {pages} {2} (\bibinfo {year}
  {2009})}\BibitemShut {NoStop}%
\bibitem [{\citenamefont {Lavina}\ \emph {et~al.}(2010)\citenamefont {Lavina},
  \citenamefont {Dera}, \citenamefont {Downs}, \citenamefont {Yang},
  \citenamefont {Sinogeikin}, \citenamefont {Meng}, \citenamefont {Shen},\ and\
  \citenamefont {Schiferl}}]{Lavina2010}%
  \BibitemOpen
  \bibfield  {author} {\bibinfo {author} {\bibfnamefont {B.}~\bibnamefont
  {Lavina}}, \bibinfo {author} {\bibfnamefont {P.}~\bibnamefont {Dera}},
  \bibinfo {author} {\bibfnamefont {R.~T.}\ \bibnamefont {Downs}}, \bibinfo
  {author} {\bibfnamefont {W.}~\bibnamefont {Yang}}, \bibinfo {author}
  {\bibfnamefont {S.}~\bibnamefont {Sinogeikin}}, \bibinfo {author}
  {\bibfnamefont {Y.}~\bibnamefont {Meng}}, \bibinfo {author} {\bibfnamefont
  {G.}~\bibnamefont {Shen}}, \ and\ \bibinfo {author} {\bibfnamefont
  {D.}~\bibnamefont {Schiferl}},\ }\href {\doibase 10.1103/PhysRevB.82.064110}
  {\bibfield  {journal} {\bibinfo  {journal} {Phys. Rev. B}\ }\textbf {\bibinfo
  {volume} {82}},\ \bibinfo {pages} {1} (\bibinfo {year} {2010})}\BibitemShut
  {NoStop}%
\bibitem [{\citenamefont {Nagai}\ \emph {et~al.}(2010)\citenamefont {Nagai},
  \citenamefont {Ishido}, \citenamefont {Seto}, \citenamefont {Nishio-Hamane},
  \citenamefont {Sata},\ and\ \citenamefont {Fujino}}]{Nagai2010}%
  \BibitemOpen
  \bibfield  {author} {\bibinfo {author} {\bibfnamefont {T.}~\bibnamefont
  {Nagai}}, \bibinfo {author} {\bibfnamefont {T.}~\bibnamefont {Ishido}},
  \bibinfo {author} {\bibfnamefont {Y.}~\bibnamefont {Seto}}, \bibinfo {author}
  {\bibfnamefont {D.}~\bibnamefont {Nishio-Hamane}}, \bibinfo {author}
  {\bibfnamefont {N.}~\bibnamefont {Sata}}, \ and\ \bibinfo {author}
  {\bibfnamefont {K.}~\bibnamefont {Fujino}},\ }\href {\doibase
  10.1088/1742-6596/215/1/012002} {\bibfield  {journal} {\bibinfo  {journal}
  {J. Phys. Conf. Ser.}\ }\textbf {\bibinfo {volume} {215}},\ \bibinfo {pages}
  {012002} (\bibinfo {year} {2010})}\BibitemShut {NoStop}%
\bibitem [{\citenamefont {Farfan}\ \emph {et~al.}(2012)\citenamefont {Farfan},
  \citenamefont {Wang}, \citenamefont {Ma}, \citenamefont {Caracas},\ and\
  \citenamefont {Mao}}]{Farfan2012}%
  \BibitemOpen
  \bibfield  {author} {\bibinfo {author} {\bibfnamefont {G.}~\bibnamefont
  {Farfan}}, \bibinfo {author} {\bibfnamefont {S.}~\bibnamefont {Wang}},
  \bibinfo {author} {\bibfnamefont {H.}~\bibnamefont {Ma}}, \bibinfo {author}
  {\bibfnamefont {R.}~\bibnamefont {Caracas}}, \ and\ \bibinfo {author}
  {\bibfnamefont {W.~L.}\ \bibnamefont {Mao}},\ }\href {\doibase
  10.2138/am.2012.4001} {\bibfield  {journal} {\bibinfo  {journal} {Am.
  Mineral.}\ }\textbf {\bibinfo {volume} {97}},\ \bibinfo {pages} {1421}
  (\bibinfo {year} {2012})}\BibitemShut {NoStop}%
\bibitem [{\citenamefont {Lin}\ \emph {et~al.}(2012)\citenamefont {Lin},
  \citenamefont {Liu}, \citenamefont {Jacobs},\ and\ \citenamefont
  {Prakapenka}}]{Lin2012}%
  \BibitemOpen
  \bibfield  {author} {\bibinfo {author} {\bibfnamefont {J.~F.}\ \bibnamefont
  {Lin}}, \bibinfo {author} {\bibfnamefont {J.}~\bibnamefont {Liu}}, \bibinfo
  {author} {\bibfnamefont {C.}~\bibnamefont {Jacobs}}, \ and\ \bibinfo {author}
  {\bibfnamefont {V.~B.}\ \bibnamefont {Prakapenka}},\ }\href {\doibase
  10.2138/am.2012.3961} {\bibfield  {journal} {\bibinfo  {journal} {Am.
  Mineral.}\ }\textbf {\bibinfo {volume} {97}},\ \bibinfo {pages} {583}
  (\bibinfo {year} {2012})}\BibitemShut {NoStop}%
\bibitem [{\citenamefont {Liu}\ \emph {et~al.}(2014)\citenamefont {Liu},
  \citenamefont {Lin}, \citenamefont {Mao},\ and\ \citenamefont
  {Prakapenka}}]{Liu2014}%
  \BibitemOpen
  \bibfield  {author} {\bibinfo {author} {\bibfnamefont {J.}~\bibnamefont
  {Liu}}, \bibinfo {author} {\bibfnamefont {J.-f.}\ \bibnamefont {Lin}},
  \bibinfo {author} {\bibfnamefont {Z.}~\bibnamefont {Mao}}, \ and\ \bibinfo
  {author} {\bibfnamefont {V.}~\bibnamefont {Prakapenka}},\ }\href@noop {}
  {\bibfield  {journal} {\bibinfo  {journal} {Am. Mineral}\ }\textbf {\bibinfo
  {volume} {99}},\ \bibinfo {pages} {84} (\bibinfo {year} {2014})}\BibitemShut
  {NoStop}%
\bibitem [{\citenamefont {Lobanov}\ \emph {et~al.}(2015)\citenamefont
  {Lobanov}, \citenamefont {Goncharov},\ and\ \citenamefont
  {Litasov}}]{Lobanov2015}%
  \BibitemOpen
  \bibfield  {author} {\bibinfo {author} {\bibfnamefont {S.~S.}\ \bibnamefont
  {Lobanov}}, \bibinfo {author} {\bibfnamefont {A.~F.}\ \bibnamefont
  {Goncharov}}, \ and\ \bibinfo {author} {\bibfnamefont {K.~D.}\ \bibnamefont
  {Litasov}},\ }\href {\doibase 10.2138/am-2015-5053} {\bibfield  {journal}
  {\bibinfo  {journal} {Am. Mineral.}\ }\textbf {\bibinfo {volume} {100}},\
  \bibinfo {pages} {1059} (\bibinfo {year} {2015})}\BibitemShut {NoStop}%
\bibitem [{\citenamefont {Hsu}\ and\ \citenamefont {Huang}(2016)}]{Hsu2016}%
  \BibitemOpen
  \bibfield  {author} {\bibinfo {author} {\bibfnamefont {H.}~\bibnamefont
  {Hsu}}\ and\ \bibinfo {author} {\bibfnamefont {S.~C.}\ \bibnamefont
  {Huang}},\ }\href {\doibase 10.1103/PhysRevB.94.060404} {\bibfield  {journal}
  {\bibinfo  {journal} {Phys. Rev. B}\ }\textbf {\bibinfo {volume} {94}},\
  \bibinfo {pages} {1} (\bibinfo {year} {2016})}\BibitemShut {NoStop}%
\bibitem [{\citenamefont {M{\"{u}}ller}\ \emph {et~al.}(2016)\citenamefont
  {M{\"{u}}ller}, \citenamefont {Speziale}, \citenamefont {Efthimiopoulos},
  \citenamefont {Jahn},\ and\ \citenamefont {Koch-M{\"{u}}ller}}]{Muller2016}%
  \BibitemOpen
  \bibfield  {author} {\bibinfo {author} {\bibfnamefont {J.}~\bibnamefont
  {M{\"{u}}ller}}, \bibinfo {author} {\bibfnamefont {S.}~\bibnamefont
  {Speziale}}, \bibinfo {author} {\bibfnamefont {I.}~\bibnamefont
  {Efthimiopoulos}}, \bibinfo {author} {\bibfnamefont {S.~A.}\ \bibnamefont
  {Jahn}}, \ and\ \bibinfo {author} {\bibfnamefont {M.}~\bibnamefont
  {Koch-M{\"{u}}ller}},\ }\href {\doibase 10.2138/am-2016-5708} {\bibfield
  {journal} {\bibinfo  {journal} {Am. Mineral.}\ }\textbf {\bibinfo {volume}
  {101}},\ \bibinfo {pages} {2638} (\bibinfo {year} {2016})}\BibitemShut
  {NoStop}%
\bibitem [{\citenamefont {M{\"{u}}ller}\ \emph {et~al.}(2017)\citenamefont
  {M{\"{u}}ller}, \citenamefont {Efthimiopoulos}, \citenamefont {Jahn},\ and\
  \citenamefont {Koch-M{\"{u}}ller}}]{Muller2017}%
  \BibitemOpen
  \bibfield  {author} {\bibinfo {author} {\bibfnamefont {J.}~\bibnamefont
  {M{\"{u}}ller}}, \bibinfo {author} {\bibfnamefont {I.}~\bibnamefont
  {Efthimiopoulos}}, \bibinfo {author} {\bibfnamefont {S.~A.}\ \bibnamefont
  {Jahn}}, \ and\ \bibinfo {author} {\bibfnamefont {M.}~\bibnamefont
  {Koch-M{\"{u}}ller}},\ }\href {\doibase 10.1127/ejm/2017/0029-2645}
  {\bibfield  {journal} {\bibinfo  {journal} {Eur. J. Mineral.}\ }\textbf
  {\bibinfo {volume} {29}},\ \bibinfo {pages} {1} (\bibinfo {year}
  {2017})}\BibitemShut {NoStop}%
\bibitem [{\citenamefont {Lin}\ \emph {et~al.}(2013)\citenamefont {Lin},
  \citenamefont {Speziale}, \citenamefont {Mao},\ and\ \citenamefont
  {Marquardt}}]{Lin2013}%
  \BibitemOpen
  \bibfield  {author} {\bibinfo {author} {\bibfnamefont {J.-f.}\ \bibnamefont
  {Lin}}, \bibinfo {author} {\bibfnamefont {S.}~\bibnamefont {Speziale}},
  \bibinfo {author} {\bibfnamefont {Z.}~\bibnamefont {Mao}}, \ and\ \bibinfo
  {author} {\bibfnamefont {H.}~\bibnamefont {Marquardt}},\ }\href {\doibase
  10.1002/rog.20010.1.INTRODUCTION} {\bibfield  {journal} {\bibinfo  {journal}
  {Rev. of Geophys.}\ }\textbf {\bibinfo {volume} {51}},\ \bibinfo {pages}
  {244} (\bibinfo {year} {2013})}\BibitemShut {NoStop}%
\bibitem [{\citenamefont {Momma}\ and\ \citenamefont {Izumi}(2008)}]{vesta}%
  \BibitemOpen
  \bibfield  {author} {\bibinfo {author} {\bibfnamefont {K.}~\bibnamefont
  {Momma}}\ and\ \bibinfo {author} {\bibfnamefont {F.}~\bibnamefont {Izumi}},\
  }\href@noop {} {\bibfield  {journal} {\bibinfo  {journal} {J. Appl. Cryst.}\
  }\textbf {\bibinfo {volume} {41}},\ \bibinfo {pages} {653} (\bibinfo {year}
  {2008})}\BibitemShut {NoStop}%
\bibitem [{\citenamefont {Hohenberg}\ and\ \citenamefont {Kohn}(1964)}]{dft1}%
  \BibitemOpen
  \bibfield  {author} {\bibinfo {author} {\bibfnamefont {P.}~\bibnamefont
  {Hohenberg}}\ and\ \bibinfo {author} {\bibfnamefont {W.}~\bibnamefont
  {Kohn}},\ }\href@noop {} {\bibfield  {journal} {\bibinfo  {journal} {Phys.\
  Rev. B}\ }\textbf {\bibinfo {volume} {136}},\ \bibinfo {pages} {864}
  (\bibinfo {year} {1964})}\BibitemShut {NoStop}%
\bibitem [{\citenamefont {Kohn}\ and\ \citenamefont {Sham}(1964)}]{dft2}%
  \BibitemOpen
  \bibfield  {author} {\bibinfo {author} {\bibfnamefont {W.}~\bibnamefont
  {Kohn}}\ and\ \bibinfo {author} {\bibfnamefont {L.~J.}\ \bibnamefont
  {Sham}},\ }\href@noop {} {\bibfield  {journal} {\bibinfo  {journal} {Phys.\
  Rev. A}\ }\textbf {\bibinfo {volume} {140}},\ \bibinfo {pages} {1133}
  (\bibinfo {year} {1964})}\BibitemShut {NoStop}%
\bibitem [{\citenamefont {Lin}\ and\ \citenamefont {Zeng}(2011)}]{Lin2011}%
  \BibitemOpen
  \bibfield  {author} {\bibinfo {author} {\bibfnamefont {H.}~\bibnamefont
  {Lin}}\ and\ \bibinfo {author} {\bibfnamefont {Z.}~\bibnamefont {Zeng}},\
  }\href@noop {} {\bibfield  {journal} {\bibinfo  {journal} {IEEE}\ }\textbf
  {\bibinfo {volume} {47}},\ \bibinfo {pages} {3817} (\bibinfo {year}
  {2011})}\BibitemShut {NoStop}%
\bibitem [{\citenamefont {Garc{\'{i}}a-Moreno}\ \emph
  {et~al.}(2001)\citenamefont {Garc{\'{i}}a-Moreno}, \citenamefont
  {Alvarez-Vega}, \citenamefont {Garc{\'{i}}a-Alvarado}, \citenamefont
  {Garc{\'{i}}a-Jaca}, \citenamefont {Gallardo-Amores}, \citenamefont
  {Sanju{\'{a}}n},\ and\ \citenamefont {Amador}}]{Garcia2001}%
  \BibitemOpen
  \bibfield  {author} {\bibinfo {author} {\bibfnamefont {O.}~\bibnamefont
  {Garc{\'{i}}a-Moreno}}, \bibinfo {author} {\bibfnamefont {M.}~\bibnamefont
  {Alvarez-Vega}}, \bibinfo {author} {\bibfnamefont {F.}~\bibnamefont
  {Garc{\'{i}}a-Alvarado}}, \bibinfo {author} {\bibfnamefont {J.}~\bibnamefont
  {Garc{\'{i}}a-Jaca}}, \bibinfo {author} {\bibfnamefont {J.~M.}\ \bibnamefont
  {Gallardo-Amores}}, \bibinfo {author} {\bibfnamefont {M.~L.}\ \bibnamefont
  {Sanju{\'{a}}n}}, \ and\ \bibinfo {author} {\bibfnamefont {U.}~\bibnamefont
  {Amador}},\ }\href@noop {} {\bibfield  {journal} {\bibinfo  {journal} {Chem.
  Mater.}\ }\textbf {\bibinfo {volume} {13}},\ \bibinfo {pages} {1570}
  (\bibinfo {year} {2001})}\BibitemShut {NoStop}%
\bibitem [{\citenamefont {Dodd}(2007)}]{Dodd2007}%
  \BibitemOpen
  \bibfield  {author} {\bibinfo {author} {\bibfnamefont {J.~L.}\ \bibnamefont
  {Dodd}},\ }\emph {\bibinfo {title} {Phase Composition and Dynamical Studies
  of Lithium Iron Phosphate}},\ \href@noop {} {Ph.D. thesis},\ \bibinfo
  {school} {California Institute of Technology}, \bibinfo {address} {Pasadena,
  California, U.S.A.} (\bibinfo {year} {2007})\BibitemShut {NoStop}%
\bibitem [{\citenamefont {Dong}\ \emph {et~al.}(2017)\citenamefont {Dong},
  \citenamefont {Guo}, \citenamefont {He}, \citenamefont {Gao}, \citenamefont
  {Han}, \citenamefont {Lu}, \citenamefont {Yan}, \citenamefont {Yang},
  \citenamefont {Li}, \citenamefont {Chen},\ and\ \citenamefont
  {Li}}]{Dong2017}%
  \BibitemOpen
  \bibfield  {author} {\bibinfo {author} {\bibfnamefont {H.}~\bibnamefont
  {Dong}}, \bibinfo {author} {\bibfnamefont {H.}~\bibnamefont {Guo}}, \bibinfo
  {author} {\bibfnamefont {Y.}~\bibnamefont {He}}, \bibinfo {author}
  {\bibfnamefont {J.}~\bibnamefont {Gao}}, \bibinfo {author} {\bibfnamefont
  {W.}~\bibnamefont {Han}}, \bibinfo {author} {\bibfnamefont {X.}~\bibnamefont
  {Lu}}, \bibinfo {author} {\bibfnamefont {S.}~\bibnamefont {Yan}}, \bibinfo
  {author} {\bibfnamefont {K.}~\bibnamefont {Yang}}, \bibinfo {author}
  {\bibfnamefont {H.}~\bibnamefont {Li}}, \bibinfo {author} {\bibfnamefont
  {D.}~\bibnamefont {Chen}}, \ and\ \bibinfo {author} {\bibfnamefont
  {H.}~\bibnamefont {Li}},\ }\href {\doibase 10.1016/j.ssi.2017.01.026}
  {\bibfield  {journal} {\bibinfo  {journal} {Solid State Ionics}\ }\textbf
  {\bibinfo {volume} {301}},\ \bibinfo {pages} {133} (\bibinfo {year}
  {2017})}\BibitemShut {NoStop}%
\bibitem [{\citenamefont {Wang}\ and\ \citenamefont {Sun}(2015)}]{Wang2015}%
  \BibitemOpen
  \bibfield  {author} {\bibinfo {author} {\bibfnamefont {J.}~\bibnamefont
  {Wang}}\ and\ \bibinfo {author} {\bibfnamefont {X.}~\bibnamefont {Sun}},\
  }\href@noop {} {\bibfield  {journal} {\bibinfo  {journal} {Energy Environ.
  Sci.}\ }\textbf {\bibinfo {volume} {8}},\ \bibinfo {pages} {1110} (\bibinfo
  {year} {2015})}\BibitemShut {NoStop}%
\bibitem [{\citenamefont {Koch-M{\"{u}}ller}\ \emph {et~al.}(2009)\citenamefont
  {Koch-M{\"{u}}ller}, \citenamefont {Mugnaioli}, \citenamefont {Rhede},
  \citenamefont {Speziale}, \citenamefont {Kolb},\ and\ \citenamefont
  {Wirth}}]{Koch-MullerMugnaioliRhedeEtAl2009}%
  \BibitemOpen
  \bibfield  {author} {\bibinfo {author} {\bibfnamefont {M.}~\bibnamefont
  {Koch-M{\"{u}}ller}}, \bibinfo {author} {\bibfnamefont {E.}~\bibnamefont
  {Mugnaioli}}, \bibinfo {author} {\bibfnamefont {D.}~\bibnamefont {Rhede}},
  \bibinfo {author} {\bibfnamefont {S.}~\bibnamefont {Speziale}}, \bibinfo
  {author} {\bibfnamefont {U.}~\bibnamefont {Kolb}}, \ and\ \bibinfo {author}
  {\bibfnamefont {R.}~\bibnamefont {Wirth}},\ }\href@noop {} {\bibfield
  {journal} {\bibinfo  {journal} {Amer. Mineral.}\ }\textbf {\bibinfo {volume}
  {99}},\ \bibinfo {pages} {2405} (\bibinfo {year} {2009})}\BibitemShut
  {NoStop}%
\bibitem [{\citenamefont {Julien}\ \emph {et~al.}(2012)\citenamefont {Julien},
  \citenamefont {Zaghib}, \citenamefont {Mauger},\ and\ \citenamefont
  {Groult}}]{JulienZaghibMaugerEtAl2012}%
  \BibitemOpen
  \bibfield  {author} {\bibinfo {author} {\bibfnamefont {C.~M.}\ \bibnamefont
  {Julien}}, \bibinfo {author} {\bibfnamefont {K.}~\bibnamefont {Zaghib}},
  \bibinfo {author} {\bibfnamefont {A.}~\bibnamefont {Mauger}}, \ and\ \bibinfo
  {author} {\bibfnamefont {H.}~\bibnamefont {Groult}},\ }\href@noop {}
  {\bibfield  {journal} {\bibinfo  {journal} {Adv. Chem. Engin. Sci.}\ }\textbf
  {\bibinfo {volume} {2}},\ \bibinfo {pages} {321} (\bibinfo {year}
  {2012})}\BibitemShut {NoStop}%
\bibitem [{\citenamefont {Syassen}(2008)}]{Syassen2008}%
  \BibitemOpen
  \bibfield  {author} {\bibinfo {author} {\bibfnamefont {K.}~\bibnamefont
  {Syassen}},\ }\href@noop {} {\bibfield  {journal} {\bibinfo  {journal} {High
  Press. Res.}\ }\textbf {\bibinfo {volume} {28}},\ \bibinfo {pages} {75}
  (\bibinfo {year} {2008})}\BibitemShut {NoStop}%
\bibitem [{\citenamefont {Mrosko}\ \emph {et~al.}(2011)\citenamefont {Mrosko},
  \citenamefont {Koch-M{\"{u}}ller},\ and\ \citenamefont
  {Schade}}]{MroskoKoch-MuellerSchade2011}%
  \BibitemOpen
  \bibfield  {author} {\bibinfo {author} {\bibfnamefont {M.}~\bibnamefont
  {Mrosko}}, \bibinfo {author} {\bibfnamefont {M.}~\bibnamefont
  {Koch-M{\"{u}}ller}}, \ and\ \bibinfo {author} {\bibfnamefont
  {U.}~\bibnamefont {Schade}},\ }\href@noop {} {\bibfield  {journal} {\bibinfo
  {journal} {Am. Mineral.}\ }\textbf {\bibinfo {volume} {96}},\ \bibinfo
  {pages} {1748} (\bibinfo {year} {2011})}\BibitemShut {NoStop}%
\bibitem [{\citenamefont {Liermann}\ \emph {et~al.}(2015)\citenamefont
  {Liermann}, \citenamefont {Konopkova}, \citenamefont {Morgenroth},
  \citenamefont {Glazyrin}, \citenamefont {Bednarcik}, \citenamefont {McBride},
  \citenamefont {Petitgirard}, \citenamefont {Delitz}, \citenamefont {Wendt},
  \citenamefont {Bican}, \citenamefont {Ehnes}, \citenamefont {Schwark},
  \citenamefont {Rothkirch}, \citenamefont {Tischer}, \citenamefont {Heuer},
  \citenamefont {Schulte-Schrepping}, \citenamefont {Kracht},\ and\
  \citenamefont {Franz}}]{LiermannKonopkovaMorgenrothEtAl2015}%
  \BibitemOpen
  \bibfield  {author} {\bibinfo {author} {\bibfnamefont {H.-P.}\ \bibnamefont
  {Liermann}}, \bibinfo {author} {\bibfnamefont {Z.}~\bibnamefont {Konopkova}},
  \bibinfo {author} {\bibfnamefont {W.}~\bibnamefont {Morgenroth}}, \bibinfo
  {author} {\bibfnamefont {K.}~\bibnamefont {Glazyrin}}, \bibinfo {author}
  {\bibfnamefont {J.}~\bibnamefont {Bednarcik}}, \bibinfo {author}
  {\bibfnamefont {E.~E.}\ \bibnamefont {McBride}}, \bibinfo {author}
  {\bibfnamefont {S.}~\bibnamefont {Petitgirard}}, \bibinfo {author}
  {\bibfnamefont {J.~T.}\ \bibnamefont {Delitz}}, \bibinfo {author}
  {\bibfnamefont {M.}~\bibnamefont {Wendt}}, \bibinfo {author} {\bibfnamefont
  {Y.}~\bibnamefont {Bican}}, \bibinfo {author} {\bibfnamefont
  {A.}~\bibnamefont {Ehnes}}, \bibinfo {author} {\bibfnamefont
  {I.}~\bibnamefont {Schwark}}, \bibinfo {author} {\bibfnamefont
  {A.}~\bibnamefont {Rothkirch}}, \bibinfo {author} {\bibfnamefont
  {M.}~\bibnamefont {Tischer}}, \bibinfo {author} {\bibfnamefont
  {J.}~\bibnamefont {Heuer}}, \bibinfo {author} {\bibfnamefont
  {H.}~\bibnamefont {Schulte-Schrepping}}, \bibinfo {author} {\bibfnamefont
  {T.}~\bibnamefont {Kracht}}, \ and\ \bibinfo {author} {\bibfnamefont
  {H.}~\bibnamefont {Franz}},\ }\href@noop {} {\bibfield  {journal} {\bibinfo
  {journal} {J. Synchrotron Rad.}\ }\textbf {\bibinfo {volume} {22}},\ \bibinfo
  {pages} {908} (\bibinfo {year} {2015})}\BibitemShut {NoStop}%
\bibitem [{\citenamefont {Hammersley}\ \emph {et~al.}(1996)\citenamefont
  {Hammersley}, \citenamefont {Svensson}, \citenamefont {Hanfland},
  \citenamefont {Fitch},\ and\ \citenamefont {Hausermann}}]{Fit2D}%
  \BibitemOpen
  \bibfield  {author} {\bibinfo {author} {\bibfnamefont {A.~P.}\ \bibnamefont
  {Hammersley}}, \bibinfo {author} {\bibfnamefont {S.~O.}\ \bibnamefont
  {Svensson}}, \bibinfo {author} {\bibfnamefont {M.}~\bibnamefont {Hanfland}},
  \bibinfo {author} {\bibfnamefont {A.~N.}\ \bibnamefont {Fitch}}, \ and\
  \bibinfo {author} {\bibfnamefont {D.}~\bibnamefont {Hausermann}},\
  }\href@noop {} {\bibfield  {journal} {\bibinfo  {journal} {High Pres. Res.}\
  }\textbf {\bibinfo {volume} {14}},\ \bibinfo {pages} {235} (\bibinfo {year}
  {1996})}\BibitemShut {NoStop}%
\bibitem [{\citenamefont {Toby}(2001)}]{Toby2001}%
  \BibitemOpen
  \bibfield  {author} {\bibinfo {author} {\bibfnamefont {B.~H.}\ \bibnamefont
  {Toby}},\ }\href@noop {} {\bibfield  {journal} {\bibinfo  {journal} {J. Appl.
  Crystallogr.}\ }\textbf {\bibinfo {volume} {34}},\ \bibinfo {pages} {210}
  (\bibinfo {year} {2001})}\BibitemShut {NoStop}%
\bibitem [{\citenamefont {Bl{\"{o}}chl}(1994)}]{blochl}%
  \BibitemOpen
  \bibfield  {author} {\bibinfo {author} {\bibfnamefont {P.~E.}\ \bibnamefont
  {Bl{\"{o}}chl}},\ }\href@noop {} {\bibfield  {journal} {\bibinfo  {journal}
  {Phys. Rev. B}\ }\textbf {\bibinfo {volume} {50}},\ \bibinfo {pages} {17953}
  (\bibinfo {year} {1994})}\BibitemShut {NoStop}%
\bibitem [{\citenamefont {Kresse}\ and\ \citenamefont
  {Joubert}(1999)}]{kresse99}%
  \BibitemOpen
  \bibfield  {author} {\bibinfo {author} {\bibfnamefont {G.}~\bibnamefont
  {Kresse}}\ and\ \bibinfo {author} {\bibfnamefont {D.}~\bibnamefont
  {Joubert}},\ }\href@noop {} {\bibfield  {journal} {\bibinfo  {journal}
  {Phys.\ Rev.\ B}\ }\textbf {\bibinfo {volume} {59}},\ \bibinfo {pages} {1758}
  (\bibinfo {year} {1999})}\BibitemShut {NoStop}%
\bibitem [{\citenamefont {Kresse}\ and\ \citenamefont
  {Furthm{\"{u}}ller}(1996{\natexlab{a}})}]{kresse96a}%
  \BibitemOpen
  \bibfield  {author} {\bibinfo {author} {\bibfnamefont {G.}~\bibnamefont
  {Kresse}}\ and\ \bibinfo {author} {\bibfnamefont {J.}~\bibnamefont
  {Furthm{\"{u}}ller}},\ }\href@noop {} {\bibfield  {journal} {\bibinfo
  {journal} {Comp. Mater. Sci.}\ }\textbf {\bibinfo {volume} {6}},\ \bibinfo
  {pages} {15} (\bibinfo {year} {1996}{\natexlab{a}})}\BibitemShut {NoStop}%
\bibitem [{\citenamefont {Kresse}\ and\ \citenamefont
  {Furthm{\"{u}}ller}(1996{\natexlab{b}})}]{kresse96b}%
  \BibitemOpen
  \bibfield  {author} {\bibinfo {author} {\bibfnamefont {G.}~\bibnamefont
  {Kresse}}\ and\ \bibinfo {author} {\bibfnamefont {J.}~\bibnamefont
  {Furthm{\"{u}}ller}},\ }\href@noop {} {\bibfield  {journal} {\bibinfo
  {journal} {Phys.\ Rev.\ B}\ }\textbf {\bibinfo {volume} {54}},\ \bibinfo
  {pages} {11169} (\bibinfo {year} {1996}{\natexlab{b}})}\BibitemShut {NoStop}%
\bibitem [{\citenamefont {{J\"{u}lich Supercomputing Centre}}(2016)}]{jureca}%
  \BibitemOpen
  \bibfield  {author} {\bibinfo {author} {\bibnamefont {{J\"{u}lich
  Supercomputing Centre}}},\ }\href {\doibase 10.17815/jlsrf-2-121} {\bibfield
  {journal} {\bibinfo  {journal} {Journal of large-scale research facilities}\
  }\textbf {\bibinfo {volume} {2}} (\bibinfo {year} {2016}),\
  10.17815/jlsrf-2-121}\BibitemShut {NoStop}%
\bibitem [{\citenamefont {Perdew}\ \emph {et~al.}(1996)\citenamefont {Perdew},
  \citenamefont {Burke},\ and\ \citenamefont {Ernzerhof}}]{perdew2}%
  \BibitemOpen
  \bibfield  {author} {\bibinfo {author} {\bibfnamefont {J.~P.}\ \bibnamefont
  {Perdew}}, \bibinfo {author} {\bibfnamefont {K.}~\bibnamefont {Burke}}, \
  and\ \bibinfo {author} {\bibfnamefont {M.}~\bibnamefont {Ernzerhof}},\
  }\href@noop {} {\bibfield  {journal} {\bibinfo  {journal} {Phys.\ Rev.\
  Lett.}\ }\textbf {\bibinfo {volume} {77}},\ \bibinfo {pages} {3865} (\bibinfo
  {year} {1996})}\BibitemShut {NoStop}%
\bibitem [{\citenamefont {Ceperley}\ and\ \citenamefont {Alder}(1980)}]{LDA}%
  \BibitemOpen
  \bibfield  {author} {\bibinfo {author} {\bibfnamefont {D.~M.}\ \bibnamefont
  {Ceperley}}\ and\ \bibinfo {author} {\bibfnamefont {B.~J.}\ \bibnamefont
  {Alder}},\ }\href@noop {} {\bibfield  {journal} {\bibinfo  {journal} {Phys.
  Rev. Lett.}\ }\textbf {\bibinfo {volume} {45}},\ \bibinfo {pages} {566}
  (\bibinfo {year} {1980})}\BibitemShut {NoStop}%
\bibitem [{\citenamefont {Anisimov}\ \emph {et~al.}(1991)\citenamefont
  {Anisimov}, \citenamefont {Zaanen},\ and\ \citenamefont
  {Andersen}}]{Anisimov1991}%
  \BibitemOpen
  \bibfield  {author} {\bibinfo {author} {\bibfnamefont {V.~I.}\ \bibnamefont
  {Anisimov}}, \bibinfo {author} {\bibfnamefont {J.}~\bibnamefont {Zaanen}}, \
  and\ \bibinfo {author} {\bibfnamefont {O.~K.}\ \bibnamefont {Andersen}},\
  }\href@noop {} {\bibfield  {journal} {\bibinfo  {journal} {Phys. Rev. B.}\
  }\textbf {\bibinfo {volume} {44}},\ \bibinfo {pages} {943} (\bibinfo {year}
  {1991})}\BibitemShut {NoStop}%
\bibitem [{\citenamefont {Dudarev}\ \emph {et~al.}(1998)\citenamefont
  {Dudarev}, \citenamefont {Botton}, \citenamefont {Savrasov}, \citenamefont
  {Humphreys},\ and\ \citenamefont {Sutton}}]{Dudarev1998}%
  \BibitemOpen
  \bibfield  {author} {\bibinfo {author} {\bibfnamefont {S.~L.}\ \bibnamefont
  {Dudarev}}, \bibinfo {author} {\bibfnamefont {G.~A.}\ \bibnamefont {Botton}},
  \bibinfo {author} {\bibfnamefont {S.~Y.}\ \bibnamefont {Savrasov}}, \bibinfo
  {author} {\bibfnamefont {C.~J.}\ \bibnamefont {Humphreys}}, \ and\ \bibinfo
  {author} {\bibfnamefont {A.~P.}\ \bibnamefont {Sutton}},\ }\href@noop {}
  {\bibfield  {journal} {\bibinfo  {journal} {Phys. Rev. B}\ }\textbf {\bibinfo
  {volume} {57}},\ \bibinfo {pages} {1505} (\bibinfo {year}
  {1998})}\BibitemShut {NoStop}%
\bibitem [{\citenamefont {Paques-Ledent}\ and\ \citenamefont
  {Tarte}(1974)}]{Paques-LedentTarte1974}%
  \BibitemOpen
  \bibfield  {author} {\bibinfo {author} {\bibfnamefont {M.~T.}\ \bibnamefont
  {Paques-Ledent}}\ and\ \bibinfo {author} {\bibfnamefont {P.}~\bibnamefont
  {Tarte}},\ }\href@noop {} {\bibfield  {journal} {\bibinfo  {journal}
  {Spectrochim. Acta A}\ }\textbf {\bibinfo {volume} {30}},\ \bibinfo {pages}
  {673} (\bibinfo {year} {1974})}\BibitemShut {NoStop}%
\bibitem [{\citenamefont {Paques-Ledent}\ and\ \citenamefont
  {Tarte}(1973)}]{Paques-LedentTarte1973}%
  \BibitemOpen
  \bibfield  {author} {\bibinfo {author} {\bibfnamefont {M.~T.}\ \bibnamefont
  {Paques-Ledent}}\ and\ \bibinfo {author} {\bibfnamefont {P.}~\bibnamefont
  {Tarte}},\ }\href@noop {} {\bibfield  {journal} {\bibinfo  {journal}
  {Spectrochim. Acta A}\ }\textbf {\bibinfo {volume} {29}},\ \bibinfo {pages}
  {1007} (\bibinfo {year} {1973})}\BibitemShut {NoStop}%
\bibitem [{\citenamefont {Paraguassu}\ \emph {et~al.}(2005)\citenamefont
  {Paraguassu}, \citenamefont {Freire}, \citenamefont {Lemos}, \citenamefont
  {Lala}, \citenamefont {Montoro},\ and\ \citenamefont
  {Rosolen}}]{ParaguassuFreireLemosEtAl2005}%
  \BibitemOpen
  \bibfield  {author} {\bibinfo {author} {\bibfnamefont {W.}~\bibnamefont
  {Paraguassu}}, \bibinfo {author} {\bibfnamefont {P.~T.~C.}\ \bibnamefont
  {Freire}}, \bibinfo {author} {\bibfnamefont {V.}~\bibnamefont {Lemos}},
  \bibinfo {author} {\bibfnamefont {S.~M.}\ \bibnamefont {Lala}}, \bibinfo
  {author} {\bibfnamefont {L.~A.}\ \bibnamefont {Montoro}}, \ and\ \bibinfo
  {author} {\bibfnamefont {J.~M.}\ \bibnamefont {Rosolen}},\ }\href@noop {}
  {\bibfield  {journal} {\bibinfo  {journal} {J. Raman Spectr.}\ }\textbf
  {\bibinfo {volume} {36}},\ \bibinfo {pages} {213} (\bibinfo {year}
  {2005})}\BibitemShut {NoStop}%
\bibitem [{\citenamefont {Burba}\ and\ \citenamefont
  {Frech}(2004)}]{BurbaFrech2004}%
  \BibitemOpen
  \bibfield  {author} {\bibinfo {author} {\bibfnamefont {C.~M.}\ \bibnamefont
  {Burba}}\ and\ \bibinfo {author} {\bibfnamefont {R.}~\bibnamefont {Frech}},\
  }\href@noop {} {\bibfield  {journal} {\bibinfo  {journal} {J. Electrochem.
  Soc.}\ }\textbf {\bibinfo {volume} {151}},\ \bibinfo {pages} {A1032}
  (\bibinfo {year} {2004})}\BibitemShut {NoStop}%
\bibitem [{\citenamefont {Ait-Salah}\ \emph {et~al.}(2006)\citenamefont
  {Ait-Salah}, \citenamefont {Dodd}, \citenamefont {Mauger}, \citenamefont
  {Yazami}, \citenamefont {Gendron},\ and\ \citenamefont
  {Julien}}]{Ait-SalahDoddMaugerEtAl2006}%
  \BibitemOpen
  \bibfield  {author} {\bibinfo {author} {\bibfnamefont {A.}~\bibnamefont
  {Ait-Salah}}, \bibinfo {author} {\bibfnamefont {J.}~\bibnamefont {Dodd}},
  \bibinfo {author} {\bibfnamefont {A.}~\bibnamefont {Mauger}}, \bibinfo
  {author} {\bibfnamefont {R.}~\bibnamefont {Yazami}}, \bibinfo {author}
  {\bibfnamefont {F.}~\bibnamefont {Gendron}}, \ and\ \bibinfo {author}
  {\bibfnamefont {C.~M.}\ \bibnamefont {Julien}},\ }\href@noop {} {\bibfield
  {journal} {\bibinfo  {journal} {Z. Anorg. Allg. Chem.}\ }\textbf {\bibinfo
  {volume} {632}},\ \bibinfo {pages} {1598} (\bibinfo {year}
  {2006})}\BibitemShut {NoStop}%
\bibitem [{\citenamefont {Klotz}\ \emph {et~al.}(2009)\citenamefont {Klotz},
  \citenamefont {Chervin}, \citenamefont {Munsch},\ and\ \citenamefont
  {Marchand}}]{Klotz2009}%
  \BibitemOpen
  \bibfield  {author} {\bibinfo {author} {\bibfnamefont {S.}~\bibnamefont
  {Klotz}}, \bibinfo {author} {\bibfnamefont {J.-C.}\ \bibnamefont {Chervin}},
  \bibinfo {author} {\bibfnamefont {P.}~\bibnamefont {Munsch}}, \ and\ \bibinfo
  {author} {\bibfnamefont {G.~L.}\ \bibnamefont {Marchand}},\ }\href@noop {}
  {\bibfield  {journal} {\bibinfo  {journal} {J. Phys. D: Appl. Phys.}\
  }\textbf {\bibinfo {volume} {42}},\ \bibinfo {pages} {075413} (\bibinfo
  {year} {2009})}\BibitemShut {NoStop}%
\bibitem [{\citenamefont {Birch}(1947)}]{Birch1947}%
  \BibitemOpen
  \bibfield  {author} {\bibinfo {author} {\bibfnamefont {F.}~\bibnamefont
  {Birch}},\ }\href@noop {} {\bibfield  {journal} {\bibinfo  {journal} {Phys.
  Rev.}\ }\textbf {\bibinfo {volume} {71}},\ \bibinfo {pages} {809} (\bibinfo
  {year} {1947})}\BibitemShut {NoStop}%
\bibitem [{\citenamefont {Maxisch}\ and\ \citenamefont
  {Ceder}(2006)}]{MaxischCeder2006}%
  \BibitemOpen
  \bibfield  {author} {\bibinfo {author} {\bibfnamefont {T.}~\bibnamefont
  {Maxisch}}\ and\ \bibinfo {author} {\bibfnamefont {G.}~\bibnamefont
  {Ceder}},\ }\href@noop {} {\bibfield  {journal} {\bibinfo  {journal} {Phys.
  Rev. B}\ }\textbf {\bibinfo {volume} {73}},\ \bibinfo {pages} {174112}
  (\bibinfo {year} {2006})}\BibitemShut {NoStop}%
\bibitem [{\citenamefont {Zhou}\ \emph {et~al.}(2004)\citenamefont {Zhou},
  \citenamefont {Cococcioni}, \citenamefont {Marianetti}, \citenamefont
  {Morgan},\ and\ \citenamefont {G.}}]{Zhou2004}%
  \BibitemOpen
  \bibfield  {author} {\bibinfo {author} {\bibfnamefont {F.}~\bibnamefont
  {Zhou}}, \bibinfo {author} {\bibfnamefont {M.}~\bibnamefont {Cococcioni}},
  \bibinfo {author} {\bibfnamefont {C.~A.}\ \bibnamefont {Marianetti}},
  \bibinfo {author} {\bibfnamefont {D.}~\bibnamefont {Morgan}}, \ and\ \bibinfo
  {author} {\bibfnamefont {C.}~\bibnamefont {G.}},\ }\href@noop {} {\bibfield
  {journal} {\bibinfo  {journal} {Phys. Rev. B}\ }\textbf {\bibinfo {volume}
  {70}},\ \bibinfo {pages} {235121} (\bibinfo {year} {2004})}\BibitemShut
  {NoStop}%
\bibitem [{\citenamefont {Shannon}\ and\ \citenamefont
  {Prewitt}(1969)}]{Shannon1969}%
  \BibitemOpen
  \bibfield  {author} {\bibinfo {author} {\bibfnamefont {R.~D.}\ \bibnamefont
  {Shannon}}\ and\ \bibinfo {author} {\bibfnamefont {C.~T.}\ \bibnamefont
  {Prewitt}},\ }\href@noop {} {\bibfield  {journal} {\bibinfo  {journal} {Acta
  Crystallogr. B}\ }\textbf {\bibinfo {volume} {25}},\ \bibinfo {pages} {925}
  (\bibinfo {year} {1969})}\BibitemShut {NoStop}%
\end{thebibliography}%


\begin{thebibliography}{6}%
\makeatletter
\providecommand \@ifxundefined [1]{%
 \@ifx{#1\undefined}
}%
\providecommand \@ifnum [1]{%
 \ifnum #1\expandafter \@firstoftwo
 \else \expandafter \@secondoftwo
 \fi
}%
\providecommand \@ifx [1]{%
 \ifx #1\expandafter \@firstoftwo
 \else \expandafter \@secondoftwo
 \fi
}%
\providecommand \natexlab [1]{#1}%
\providecommand \enquote  [1]{``#1''}%
\providecommand \bibnamefont  [1]{#1}%
\providecommand \bibfnamefont [1]{#1}%
\providecommand \citenamefont [1]{#1}%
\providecommand \href@noop [0]{\@secondoftwo}%
\providecommand \href [0]{\begingroup \@sanitize@url \@href}%
\providecommand \@href[1]{\@@startlink{#1}\@@href}%
\providecommand \@@href[1]{\endgroup#1\@@endlink}%
\providecommand \@sanitize@url [0]{\catcode `\\12\catcode `\$12\catcode
  `\&12\catcode `\#12\catcode `\^12\catcode `\_12\catcode `\%12\relax}%
\providecommand \@@startlink[1]{}%
\providecommand \@@endlink[0]{}%
\providecommand \url  [0]{\begingroup\@sanitize@url \@url }%
\providecommand \@url [1]{\endgroup\@href {#1}{\urlprefix }}%
\providecommand \urlprefix  [0]{URL }%
\providecommand \Eprint [0]{\href }%
\providecommand \doibase [0]{http://dx.doi.org/}%
\providecommand \selectlanguage [0]{\@gobble}%
\providecommand \bibinfo  [0]{\@secondoftwo}%
\providecommand \bibfield  [0]{\@secondoftwo}%
\providecommand \translation [1]{[#1]}%
\providecommand \BibitemOpen [0]{}%
\providecommand \bibitemStop [0]{}%
\providecommand \bibitemNoStop [0]{.\EOS\space}%
\providecommand \EOS [0]{\spacefactor3000\relax}%
\providecommand \BibitemShut  [1]{\csname bibitem#1\endcsname}%
\let\auto@bib@innerbib\@empty
\bibitem [{\citenamefont {d'Yvoire}\ \emph {et~al.}(1983)\citenamefont
  {d'Yvoire}, \citenamefont {Pintard-Screpel}, \citenamefont {Bretey},\ and\
  \citenamefont {de~la Rochere}}]{Pintard-ScrepelBreteyEtAl1983}%
  \BibitemOpen
  \bibfield  {author} {\bibinfo {author} {\bibfnamefont {F.}~\bibnamefont
  {d'Yvoire}}, \bibinfo {author} {\bibfnamefont {M.}~\bibnamefont
  {Pintard-Screpel}}, \bibinfo {author} {\bibfnamefont {E.}~\bibnamefont
  {Bretey}}, \ and\ \bibinfo {author} {\bibfnamefont {M.}~\bibnamefont {de~la
  Rochere}},\ }\href@noop {} {\bibfield  {journal} {\bibinfo  {journal} {Sol.
  St. Ionics}\ }\textbf {\bibinfo {volume} {9-10}},\ \bibinfo {pages} {851}
  (\bibinfo {year} {1983})}\BibitemShut {NoStop}%
\bibitem [{\citenamefont {Bih}\ \emph {et~al.}(2009)\citenamefont {Bih},
  \citenamefont {Bih}, \citenamefont {Manoun}, \citenamefont {Azdouz},
  \citenamefont {Benmokhtar},\ and\ \citenamefont
  {Lazor}}]{BihBihManounEtAl2009}%
  \BibitemOpen
  \bibfield  {author} {\bibinfo {author} {\bibfnamefont {H.}~\bibnamefont
  {Bih}}, \bibinfo {author} {\bibfnamefont {L.}~\bibnamefont {Bih}}, \bibinfo
  {author} {\bibfnamefont {B.}~\bibnamefont {Manoun}}, \bibinfo {author}
  {\bibfnamefont {M.}~\bibnamefont {Azdouz}}, \bibinfo {author} {\bibfnamefont
  {S.}~\bibnamefont {Benmokhtar}}, \ and\ \bibinfo {author} {\bibfnamefont
  {P.}~\bibnamefont {Lazor}},\ }\href@noop {} {\bibfield  {journal} {\bibinfo
  {journal} {J. Molecul. Str.}\ }\textbf {\bibinfo {volume} {936}},\ \bibinfo
  {pages} {147} (\bibinfo {year} {2009})}\BibitemShut {NoStop}%
\bibitem [{\citenamefont {Markevich}\ \emph {et~al.}(2011)\citenamefont
  {Markevich}, \citenamefont {Sharabi}, \citenamefont {Haik}, \citenamefont
  {Borgel}, \citenamefont {Salitra}, \citenamefont {Aurbach}, \citenamefont
  {Semrau}, \citenamefont {Schmidt}, \citenamefont {Schall},\ and\
  \citenamefont {Stinner}}]{MarkevichSharabiHaikEtAl2011}%
  \BibitemOpen
  \bibfield  {author} {\bibinfo {author} {\bibfnamefont {E.}~\bibnamefont
  {Markevich}}, \bibinfo {author} {\bibfnamefont {R.}~\bibnamefont {Sharabi}},
  \bibinfo {author} {\bibfnamefont {O.}~\bibnamefont {Haik}}, \bibinfo {author}
  {\bibfnamefont {V.}~\bibnamefont {Borgel}}, \bibinfo {author} {\bibfnamefont
  {G.}~\bibnamefont {Salitra}}, \bibinfo {author} {\bibfnamefont
  {D.}~\bibnamefont {Aurbach}}, \bibinfo {author} {\bibfnamefont
  {G.}~\bibnamefont {Semrau}}, \bibinfo {author} {\bibfnamefont {M.~A.}\
  \bibnamefont {Schmidt}}, \bibinfo {author} {\bibfnamefont {N.}~\bibnamefont
  {Schall}}, \ and\ \bibinfo {author} {\bibfnamefont {C.}~\bibnamefont
  {Stinner}},\ }\href@noop {} {\bibfield  {journal} {\bibinfo  {journal} {J.
  Power Sources}\ }\textbf {\bibinfo {volume} {196}},\ \bibinfo {pages} {6433}
  (\bibinfo {year} {2011})}\BibitemShut {NoStop}%
\bibitem [{\citenamefont {Burba}\ and\ \citenamefont
  {Frech}(2004)}]{BurbaFrech2004}%
  \BibitemOpen
  \bibfield  {author} {\bibinfo {author} {\bibfnamefont {C.~M.}\ \bibnamefont
  {Burba}}\ and\ \bibinfo {author} {\bibfnamefont {R.}~\bibnamefont {Frech}},\
  }\href@noop {} {\bibfield  {journal} {\bibinfo  {journal} {J. Electrochem.
  Soc.}\ }\textbf {\bibinfo {volume} {151}},\ \bibinfo {pages} {A1032}
  (\bibinfo {year} {2004})}\BibitemShut {NoStop}%
\bibitem [{\citenamefont {Paraguassu}\ \emph {et~al.}(2005)\citenamefont
  {Paraguassu}, \citenamefont {Freire}, \citenamefont {Lemos}, \citenamefont
  {Lala}, \citenamefont {Montoro},\ and\ \citenamefont
  {Rosolen}}]{ParaguassuFreireLemosEtAl2005}%
  \BibitemOpen
  \bibfield  {author} {\bibinfo {author} {\bibfnamefont {W.}~\bibnamefont
  {Paraguassu}}, \bibinfo {author} {\bibfnamefont {P.~T.~C.}\ \bibnamefont
  {Freire}}, \bibinfo {author} {\bibfnamefont {V.}~\bibnamefont {Lemos}},
  \bibinfo {author} {\bibfnamefont {S.~M.}\ \bibnamefont {Lala}}, \bibinfo
  {author} {\bibfnamefont {L.~A.}\ \bibnamefont {Montoro}}, \ and\ \bibinfo
  {author} {\bibfnamefont {J.~M.}\ \bibnamefont {Rosolen}},\ }\href@noop {}
  {\bibfield  {journal} {\bibinfo  {journal} {J. Raman Spectr.}\ }\textbf
  {\bibinfo {volume} {36}},\ \bibinfo {pages} {213} (\bibinfo {year}
  {2005})}\BibitemShut {NoStop}%
\bibitem [{\citenamefont {Zhou}\ \emph {et~al.}(2004)\citenamefont {Zhou},
  \citenamefont {Cococcioni}, \citenamefont {Marianetti}, \citenamefont
  {Morgan},\ and\ \citenamefont {G.}}]{Zhou2004}%
  \BibitemOpen
  \bibfield  {author} {\bibinfo {author} {\bibfnamefont {F.}~\bibnamefont
  {Zhou}}, \bibinfo {author} {\bibfnamefont {M.}~\bibnamefont {Cococcioni}},
  \bibinfo {author} {\bibfnamefont {C.~A.}\ \bibnamefont {Marianetti}},
  \bibinfo {author} {\bibfnamefont {D.}~\bibnamefont {Morgan}}, \ and\ \bibinfo
  {author} {\bibfnamefont {C.}~\bibnamefont {G.}},\ }\href@noop {} {\bibfield
  {journal} {\bibinfo  {journal} {Phys. Rev. B}\ }\textbf {\bibinfo {volume}
  {70}},\ \bibinfo {pages} {235121} (\bibinfo {year} {2004})}\BibitemShut
  {NoStop}%
\end{thebibliography}%

\end{document}


\preprint{APS/123-QED}

\title{Evidence for a pressure-induced spin transition in olivine-type LiFePO$_{4}$ triphylite}

\author{Maribel N\'u\~nez Valdez}
\email{mari\_nv@gfz-potsdam.de}
\affiliation{GFZ German Research Centre for Geosciences, Telegrafenberg, 14473 Potsdam, Germany}

\author{Ilias Efthimiopoulos}
\email{iliefthi@gfz-potsdam.de}
\affiliation{GFZ German Research Centre for Geosciences, Telegrafenberg, 14473 Potsdam, Germany}
\affiliation{Institute of Earth and Environmental Science, University of Potsdam, Karl-Liebknecht-Strasse 24-25, 14476 Potsdam-Golm, Germany}%

\author{Michail Taran}
\affiliation{Natl Acad Sci Ukraine, Inst Geochem Mineral and Ore Format, UA-03680 Kiev 142, Ukraine}%

\author{Jan M\"uller}
\affiliation{GFZ German Research Centre for Geosciences, Telegrafenberg, 14473 Potsdam, Germany}
\affiliation{Institute of Geology and Mineralogy, University of Cologne, Zülpicher Str. 49b, 50674 Cologne, Germany}%

\author{Elena Bykova}
\affiliation{FS-PE, PETRA III, Deutsches Elektronen Synchrotron, 22607 Hamburg, Germany}%

\author{Monika Koch-M\"uller}
\affiliation{GFZ German Research Centre for Geosciences, Telegrafenberg, 14473 Potsdam, Germany}

\author{Max Wilke}
\affiliation{Institute of Earth and Environmental Science, University of Potsdam, Karl-Liebknecht-Strasse 24-25, 14476 Potsdam-Golm, Germany}%

\date{\today}
\begin{abstract}
$\textbf{SUPPLEMENTAL MATERIAL}$
\end{abstract}
\pacs{Valid PACS appear here}
\maketitle
\begin{figure*}
\begin{subfigure}{0.4\textwidth}
  \includegraphics[width=\textwidth]{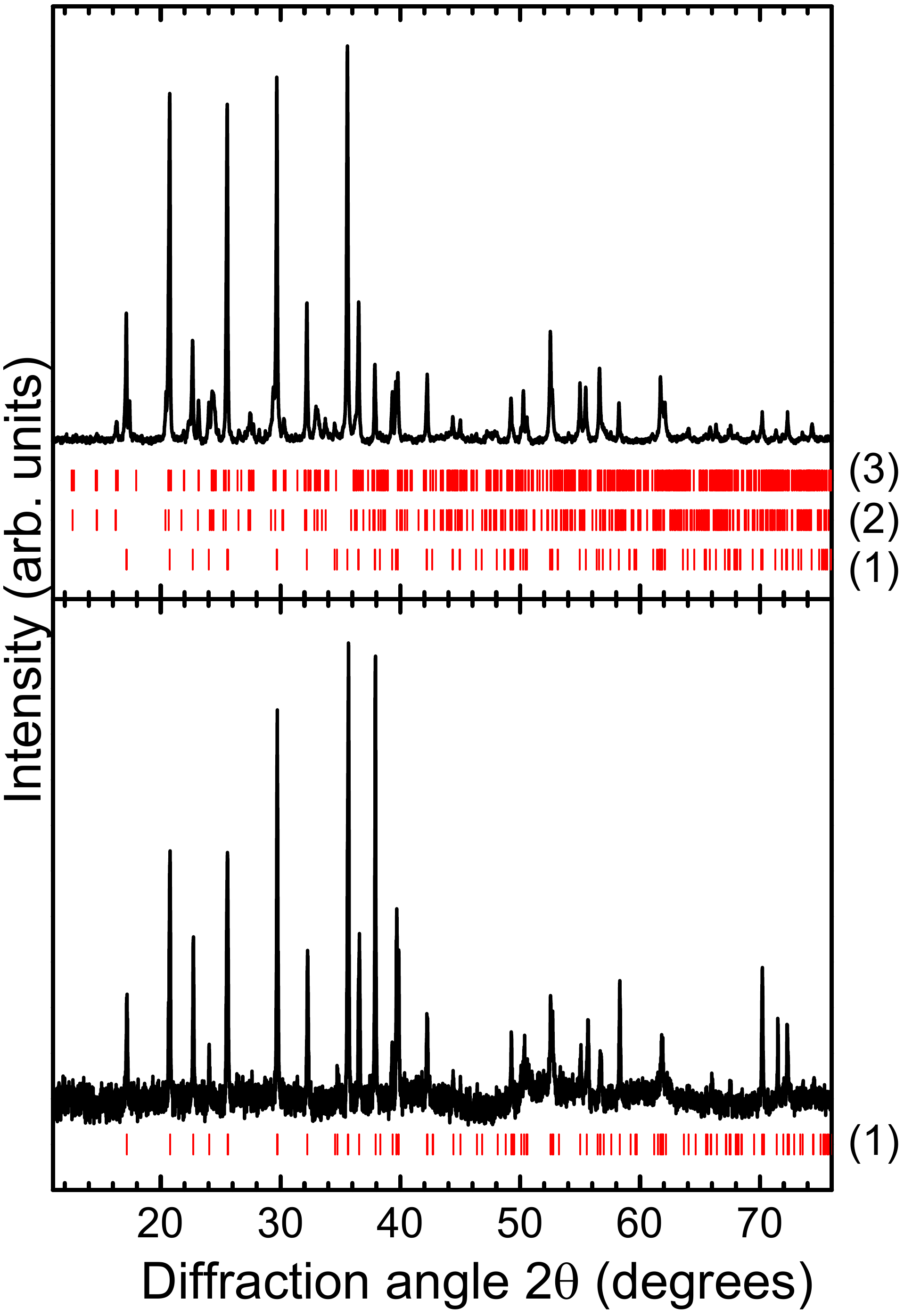}
   \caption{} \label{Fig_S1}
\end{subfigure}
\begin{subfigure}{0.4\textwidth}
        \includegraphics[width=\textwidth]{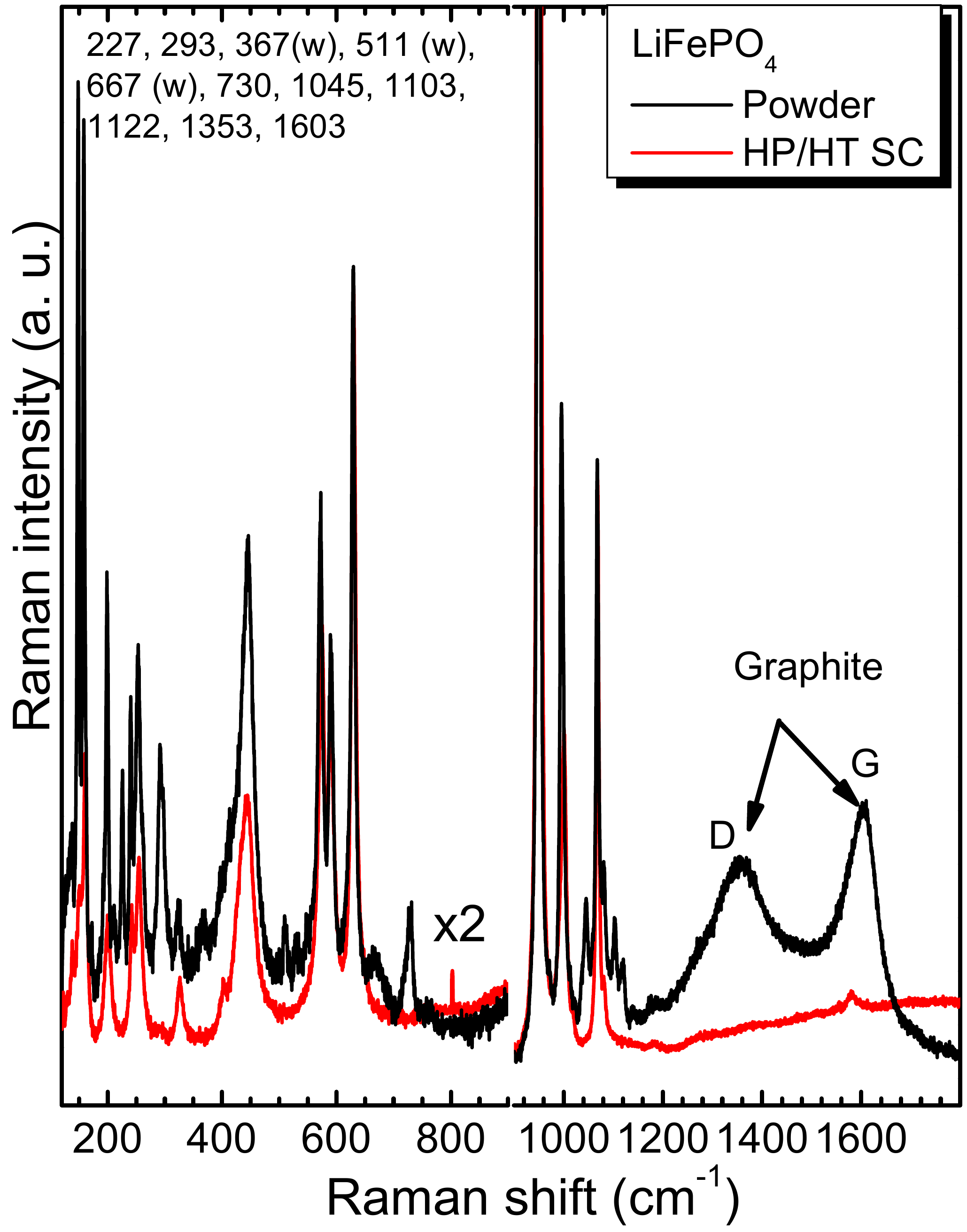}
        \caption{} \label{Fig_S2}
    \end{subfigure}
    \caption{(a) XRD patterns of the  LiFePO$_{4}$ samples used in our studies (sample A top, sample B bottom) at ambient conditions measured with a STOE Stadi P diffractometer with a primary monochromator and Cu-K$_{\alpha1}$ radiation ($\lambda_{Cu}$~=~1.05456~{\AA}). Black solid lines correspond to the measured patterns, whereas the vertical ticks mark the Bragg peak positions for the various phases. The latter are distinguished by numbers: (1) LiFePO$_{4}$ (SG $Pbnm$, $Z$ = 4), (2) Li$_{3}$Fe$_{2}$(PO$_{4}$)$_{3}$ (SG $P2_{1}$/$c$, $Z$ = 4), and (3) Li$_{3}$Fe$_{2}$(PO$_{4}$)$_{3}$ (SG $R\bar{3}$, $Z$ = 6). Lattice parameters and volume for the LiFePO$_{4}$ sample A are: $a_{0}$~=~10.3214(1)~{\AA}, $b_{0}$ = 6.0017(3)~{\AA}, $c_{0}$~=~4.7291(2)~{\AA}, $V_{0}$~=~292.95~{\AA}$^{3}$; and for LiFePO$_{4}$ sample B: $a_{0}$~=~10.3186(1)~{\AA}, $b_{0}$ = 6.0026(3)~{\AA}, $c_{0}$~=~4.6951(2)~{\AA}, $V_{0}$~=~290.81~{\AA}$^{3}$. (b) Measured Raman spectra of the two samples of LiFePO$_{4}$ (powder sample A, black, and single-crystalline sample B, red) at ambient conditions ($\lambda$~=~473~nm). The additional Raman features of sample A are listed in the upper left corner.}
\end{figure*}
\begin{figure*}
\begin{subfigure}{0.4\textwidth}
  \includegraphics[width=\textwidth]{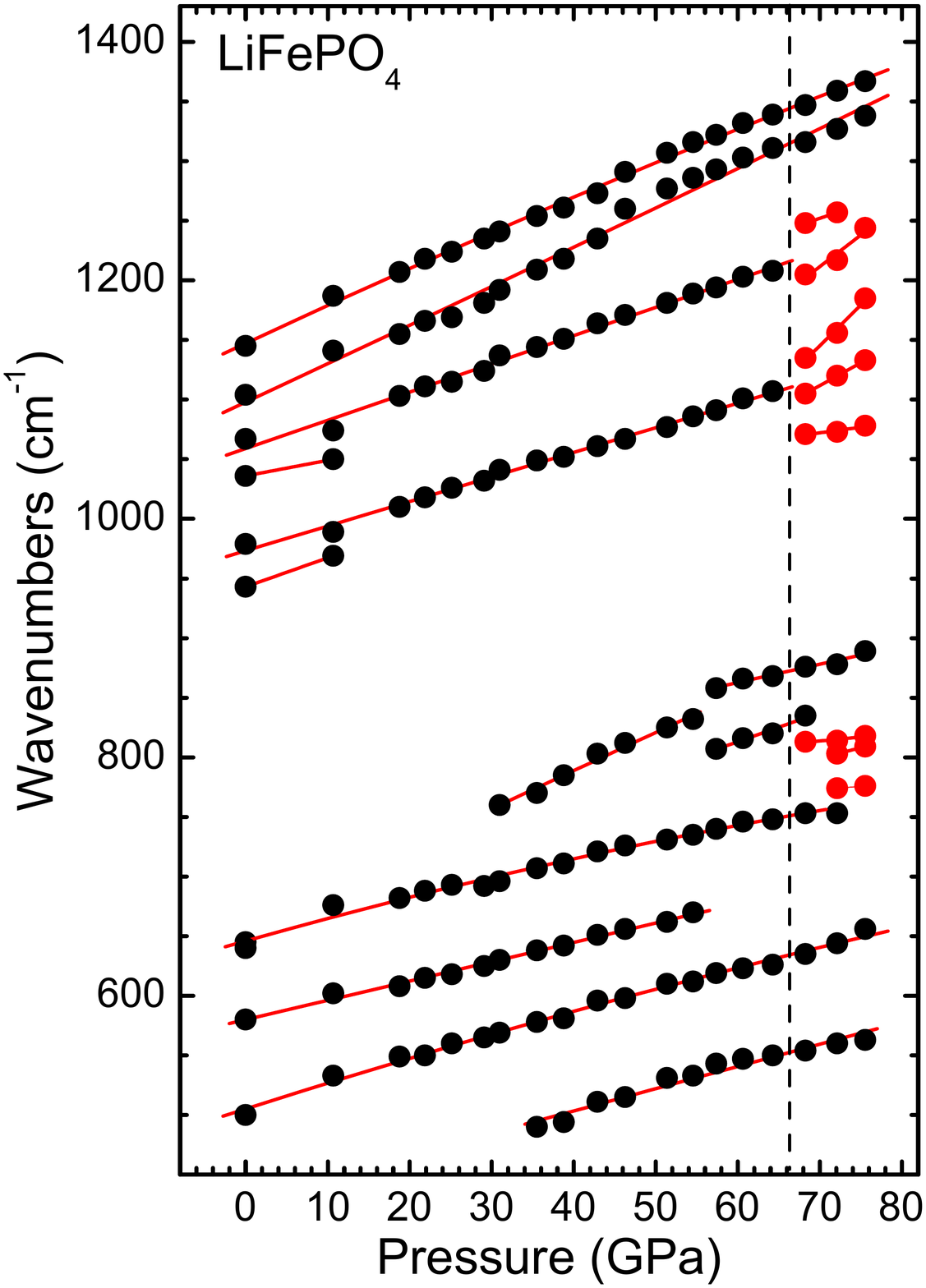}
   \caption{} \label{Fig_S3}
\end{subfigure}
\begin{subfigure}{0.4\textwidth}
        \includegraphics[width=\textwidth]{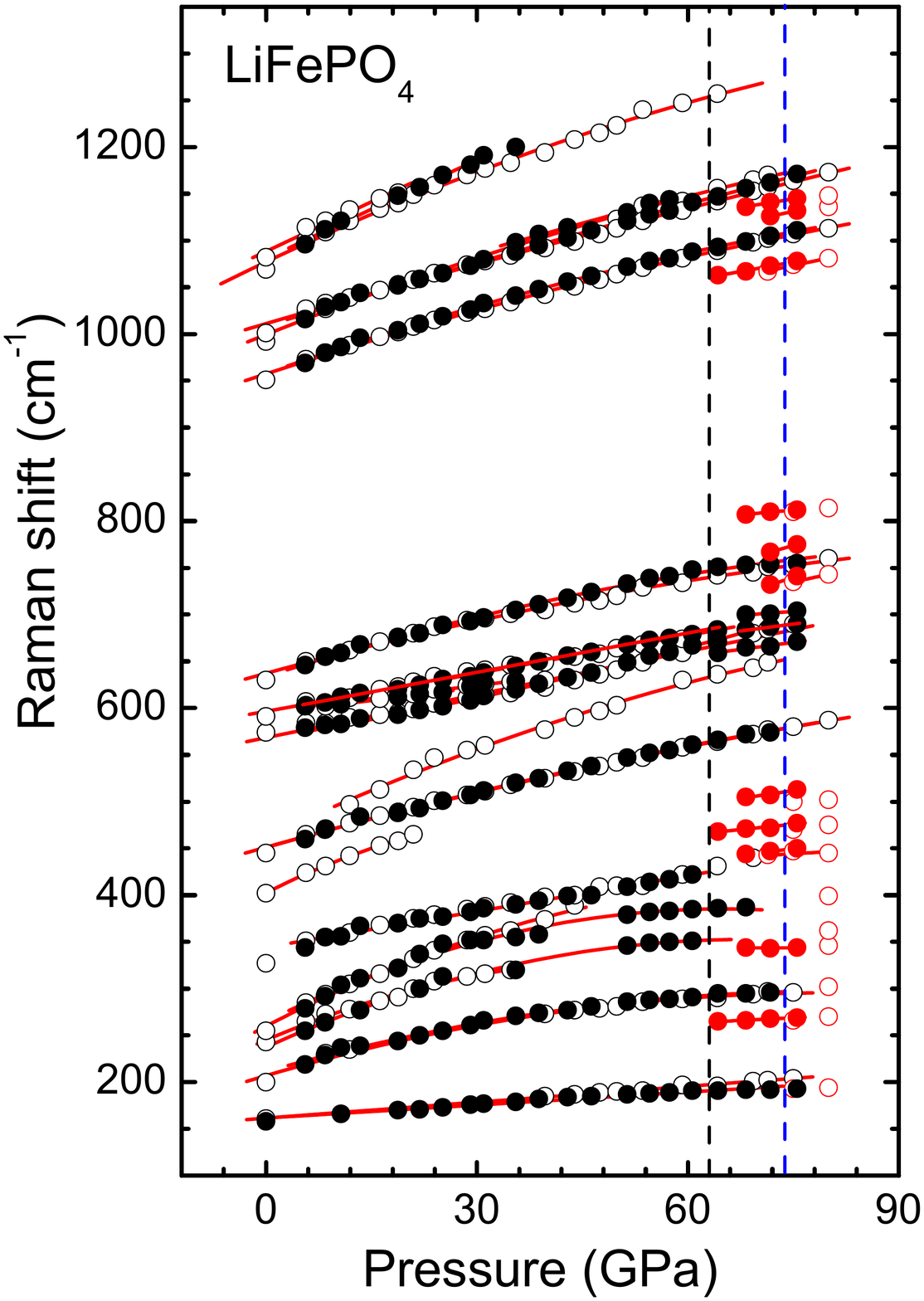}
        \caption{} \label{Fig_S4}
    \end{subfigure}
    \caption{Evolution of the (a) MIR and (b) Raman mode frequencies as a function of pressure. The closed and open symbols correspond to data collected from sample A (powder) and sample B (single-crystal), respectively. The vertical dashed lines represent the onset of the isostructural transition in each case (black for sample A, blue for sample B).}
\end{figure*}

The XRD and Raman measurements of the synthetic powder (sample A) and the HP/HT-synthesized LiFePO$_{4}$ single crystals (sample B) at ambient conditions are shown in Figs.~1(a) and Fig. 1(b), respectively. The XRD analysis of sample A revealed the presence of two additional impurities, which were identified as the monoclinic (SG 14, $P2_{1}/c$, $Z$ = 4) and rhombohedral (SG 148, $R\bar{3}$, $Z$ = 6) modifications of Li$_{3}$Fe$_{2}$(PO$_{4}$)$_{3}$ [Fig. 1(a)]. We note here that the rhombohedral Li$_{3}$Fe$_{2}$(PO$_{4}$)$_{3}$ phase is synthesized by high-temperature solid-state reaction, whereas heating of this phase above 840~K leads to the adoption of the $P2_{1}$/$c$ structure\cite{Pintard-ScrepelBreteyEtAl1983}. The latter survives upon cooling to room temperature. The estimate for the concentration of these Fe$^{3+}$-bearing impurity phases from both the Rietveld refinement of the XRD pattern [Fig. 1(a)], as well as from M\"{o}ssbauer measurements (not shown) lies at $\sim$30\%.

Our Raman spectra corroborate the XRD results. In Fig. 2 we plot the Raman response of the two LiFePO$_{4}$ samples. As we can observe, the powder sample (sample A) exhibits a higher number of Raman peaks compared to the single-crystalline one (sample B). These additional Raman bands can be assigned partly to the monoclinic  Li$_{3}$Fe$_{2}$(PO$_{4}$)$_{3}$  phase\cite{BihBihManounEtAl2009}; we could not find any Raman data on the rhombohedral Li$_{3}$Fe$_{2}$(PO$_{4}$)$_{3}$  modification. In addition, the two strong and broad Raman bands located at 1380 cm$^{-1}$ and 1600 cm$^{-1}$ can be attributed to the D and G band of graphite\cite{MarkevichSharabiHaikEtAl2011}, resulting from the carbon coating of the commercial LiFePO$_{4}$ sample. We note though that we could not detect any graphite in our XRD pattern [Fig. 1(a)], owing probably to the small $Z$ number of carbon.

\begin{table*}%
\caption{\label{Table_S1}%
  Assignment\cite{BurbaFrech2004,ParaguassuFreireLemosEtAl2005}, frequencies, pressure coefficients, and mode Gr\"{u}neisen parameters $\gamma_{i}$ of the experimentally observed MIR-active modes for the phases of the LiFePO$_{4}$ powder sample (sample A), evaluated at pressure $P$. The pressure dependence is given by the relation: $\omega(P)=\omega_{R}+aP+bP^{2}$. Gr\"{u}neisen parameters are determined from the relation:
  $\gamma_{i}=\left(\frac{B_{0}}{\omega_{Ri}}\right)\left(\frac{\partial\omega_{Ri}}{\partial P}\right)$. We have employed the bulk moduli values $B_{0}=99$ GPa for the starting $Pbnm$ phase, and $B$~=~545 GPa for the HP-$Pbnm$ modification, as measured here (see Table 1 in main text).}
\begin{ruledtabular}
\begin{tabular}{lccdddd}
 Phase &  Mode symmetry (IR)  & $P$ (GPa)& \multicolumn{1}{c}{\textrm{$\omega_{R}$ (cm$^{-1}$)}} & \multicolumn{1}{c}{\textrm{$\frac{\partial\omega_{Ri}}{\partial P}$ (cm$^{-1}$/GPa)}} &  \multicolumn{1}{c}{\textrm{$\frac{\partial^2\omega_{Ri}}{\partial P^2}$ (cm$^{-1}$/GPa$^{2}$)}} & \multicolumn{1}{c}{\textrm{$\gamma_{i}$}} \\
   \colrule
 $Pbnm$ & $\nu_{4}+\nu_{2}$  & 0 & 429 & 1.9 & - & 0.44 \\
     &$\nu_{4}+\nu_{2}$  &   & 505 & 2.2 & -0.001 & 0.43 \\
    & $\nu_{4}+\nu_{2}$  &   & 580 & 1.6 & - & 0.27 \\
    & $\nu_{4}$ &  & 646  & 1.9 & -0.005 & 0.29 \\
   & $\nu_{4}$ &   & 661  & 3.2 & - & 0.48 \\
   &$\nu_{4}$ &    & 667  & 2.4 & - & 0.36 \\
   &$\nu_{4}$ &    & 771  & 1.5 & - & 0.19 \\
       &$\nu_{1}$ &    & 943 & 2.4 & - & 0.25 \\
   &$\nu_{1}$ &    & 973  & 2.1 & - & 0.21 \\
   & $\nu_{3}$ &  & 1036  & 1.3 & - & 0.12 \\
   & $\nu_{3}$ &   & 1059  & 2.4 & - & 0.22 \\
   &$\nu_{3}$ &    & 1097  & 3.3 & - & 0.30 \\
   &$\nu_{3}$ &    & 1147  & 3.2 & -0.004 & 0.28 \\
  \colrule
 HP-$Pbnm$ & $\nu_{4}$  &73 & 776 & 0.6 & - & 0.08 \\
   & $\nu_{4}$ &  & 807 & 1.8 & - & 0.22\\
    & $\nu_{4}$ &  & 817 & 0.7 & - & 0.08\\
    & $\nu_{1}$ &  & 1079 & 1.0 & - & 0.09\\
    & $\nu_{1}$ &  & 1120 & 3.8 & - & 0.34\\
    & $\nu_{3}$ &  & 1163 & 6.8 & - & 0.58\\
    & $\nu_{3}$ &  & 1228 & 5.3 & - & 0.43\\
     & $\nu_{3}$ &  & 1256 & 2.3 & - & 0.18\\
\end{tabular}
\end{ruledtabular}
\end{table*}

\begin{table*}%
  \caption{\label{Table_S2}%
  Assignment\cite{BurbaFrech2004,ParaguassuFreireLemosEtAl2005}, frequencies, pressure coefficients, and mode Gr\"{u}neisen parameters $\gamma_{i}$ of the experimentally observed Raman-active modes for the phases of the LiFePO$_{4}$ powder sample (sample A), evaluated at pressure $P$. The pressure dependence is given by the relation: $\omega(P)=\omega_{R}+aP+bP^{2}$. Gr\"{u}neisen parameters are determined from the relation:
  $\gamma_{i}=\left(\frac{B_{0}}{\omega_{Ri}}\right)\left(\frac{\partial\omega_{Ri}}{\partial P}\right)$. We have employed the bulk moduli values $B_{0}=99$ GPa for the starting $Pbnm$ phase, and $B=545$ GPa for the HP-$Pbnm$ modification, as measured here (see Table 1 in main text). The numbers in parentheses correspond to the respective single-crystalline Raman data (sample B).}
\begin{ruledtabular}
\begin{tabular}{lcccddd}
 Phase &  Mode symmetry (IR)  & $P$ (GPa)& \multicolumn{1}{c}{\textrm{$\omega_{R}$ (cm$^{-1}$)}} & \multicolumn{1}{c}{\textrm{$\frac{\partial\omega_{Ri}}{\partial P}$ (cm$^{-1}$/GPa)}} &  \multicolumn{1}{c}{\textrm{$\frac{\partial^2\omega_{Ri}}{\partial P^2}$ (cm$^{-1}$/GPa$^{2}$)}} & \multicolumn{1}{c}{\textrm{$\gamma_{i}$}} \\
   \colrule
 $Pbnm$ & A$_{g}$-T(PO$_{4}$)+T(M)  & 0 & 162(161) & 0.5(0.6) & - & 0.31(0.37) \\
     & B$_{1g}$-L(PO$_{4}$)  &   & 211(207) & 2.1(2.2) & -0.01(-0.01) & 0.99(1.1) \\
 & B$_{1g}$-L(PO$_{4}$+T$_{w}$(M)  &   & 238(245) & 3.4(3.2) & -0.02(-0.03) & 1.41(1.29) \\
 & A$_{g}$-L(PO$_{4}$)+T(M) &  & 265(260)  & 3.7(3.9) & -0.03(-0.02) & 1.38(1.49) \\
 & A$_{g}$-L(PO$_{4}$)+T(M)&   & 344(336)  & 1.3(1.9) & (-0.01) & 0.37(0.56) \\
& A$_{g}$-T(PO$_{4}$)+T(M)  &   & (402) & (3.8) & (-0.04) & (0.94) \\
& B$_{1g}$-$\nu_{2}$  &   & 451(451) & 2.1(2.1) & -0.01(-0.01) & 0.46(0.46) \\
& B$_{3g}$-$\nu_{2}$  &   & (462) & (2.1) & (-0.01) & (0.45) \\
& A$_{g}$-$\nu_{4}$  &   & 568(568) & 1.3(1.5) & 0.01 & 0.23(0.26) \\
& B$_{1g}$-$\nu_{4}$  &   & 594(596) & 0.9(0.9) & - & 0.15(0.15) \\
& B$_{2g}$-$\nu_{4}$  &   & 595(596) & 1.3(1.4) & - & 0.22(0.23) \\
& A$_{g}$-$\nu_{4}$  &   & 634(637) & 2.4(2.1) & -0.01(-0.01) & 0.37(0.33) \\
& A$_{g}$-$\nu_{1}$  &   & 958(957) & 2.6(2.5) & -0.01(-0.01) & 0.27(0.26) \\
& B$_{1g}$-$\nu_{3}$  &   & (1000) & (3.0) & (-0.01) & (0.30) \\
& A$_{g}$-$\nu_{3}$  &   & 1008(1011) & 2.3(2.2) & -0.01(-0.01) & 0.23(0.22)\\
& B$_{2g}$-$\nu_{3}$  &   & 1025 & 2.1 & - & 0.20 \\
& A$_{g}$-$\nu_{3}$  &   & 1081(1077) & 3.5(3.6) & (-0.01) & 0.32(0.33) \\

\colrule

 HP-$Pbnm$ & B$_{1g}$-L(PO$_{4}$+T$_{w}$(M)  & 73 & 270 & 0.40 & - & 0.81 \\
   & A$_{g}$-L(PO$_{4}$)+T(M) &   & 344 & 0.01 & - & 0.02\\
   & A$_{g}$-T(PO$_{4}$)+T(M)  &   & 446(447) & 0.8(0.5) & - & 0.98(0.61)\\
   & B$_{1g}$-$\nu_{2}$   &   & 477 & 0.8 & - & 0.91\\
   & B$_{3g}$-$\nu_{2}$  &   & 509 & 1.1 & - & 1.18\\
   & B$_{1g}$-$\nu_{4}$  &   & 668 & 1.0 & - & 0.82\\
   & B$_{2g}$-$\nu_{4}$  &   & 688 & 0.5 & - & 0.40\\
   & A$_{g}$-$\nu_{4}$  &   & 706 & 0.6 & - & 0.46\\
   & A$_{g}$-$\nu_{4}$  &   & 738(732) & 2.4(1.6) & - & 1.77(1.19)\\
  & A$_{g}$-$\nu_{4}$  &   & 770 & 2.1 & - & 1.49\\
   & B$_{2g}$-$\nu_{4}$  &   & 812 & 0.7 & - & 0.47\\
   & A$_{g}$-$\nu_{3}$   &   & 1077(1070)& 1.4(1.6) & - & 0.71(0.82)\\
   & B$_{2g}$-$\nu_{3}$  &   & 1130 & 1.6 & - & 0.77\\
  & A$_{g}$-$\nu_{3}$  &   & 1140 & 1.2 & - & 0.57\\

\end{tabular}
\end{ruledtabular}
\end{table*}
\begin{table*}%
\caption{\label{Table_S2}%
Experimental lattice parameters for the $Pbnm$ and the HP-$Pbnm$ phases of LiFePO$_{4}$ (HP/HT single crystals, sample B).}
\begin{ruledtabular}
\begin{tabular}{lddddd}
 Phase &  \multicolumn{1}{c}{\textrm{$P$ (GPa)}}  & \multicolumn{1}{c}{\textrm{$a$ ({\AA})}} & \multicolumn{1}{c}{\textrm{$b$ ({\AA})}} &  \multicolumn{1}{c}{\textrm{$c$ ({\AA})}} &  \multicolumn{1}{c}{\textrm{$V$ ({\AA}$^{3}$)}} \\
   \colrule
          $Pbnm$ & 0.0  & 4.6951(2) & 10.3186(3) & 6.0026(1) & 290.8 \\
       & 3.0  & 4.635(2) & 10.156(3) & 6.001(1) & 282.4 \\
       & 15.5  & 4.588(2) & 9.777(3) & 5.901(1) & 264.7 \\
       & 19.0  & 4.560(2) & 9.691(3) & 5.848(1) & 258.5 \\
       & 22.0  & 4.531(2) & 9.654(3) & 5.818(1) & 254.5 \\
       & 24.0  & 4.506(2) & 9.618(3) & 5.796(1) & 251.2 \\
       & 28.0  & 4.464(2) & 9.551(3) & 5.772(1) & 246.0 \\
       & 31.4  & 4.447(2) & 9.485(3) & 5.747(1) & 242.4 \\
       &34.6                    &  4.411(2) & 9.356(3) & 5.702(1) & 235.3 \\
        & 38.0                  &   4.392(2)   & 9.308(3) & 5.694(1) & 232.8\\
         &  40.5         &   4.359(2)   & 9.311(3) & 5.677(1) &  230.4\\
   &43.2 & 4.355(2) & 9.281(3) &  5.533(1) & 228.1\\
    & 48.1     & 4.328(2) &  9.269(3) &  5.490(1) & 224.9 \\
   & 55.5 & 4.316(2) & 9.154(3) & 5.451(1) & 218.6\\
    & 61.5 & 4.298(2) & 9.086(3) & 5.307(1) & 214.4\\
    & 68.5 & 4.270(2) & 9.025(3) & 5.269(1) & 210.1\\
    \colrule
    HP-$Pbnm$ &73.0 & 4.256(2) & 9.001(3) & 5.307(1) & 203.3\\
   & 78.5 & 4.245(2) & 8.976(3) & 5.269(1) & 200.8\\
\end{tabular}
\end{ruledtabular}
\end{table*}

\begin{table*}%
\caption{\label{Table_S3}%
Lattice parameters and volume for the $Pbnm$-LiFePO$_{4}$ in the AFM-HS state from our GGA+U calculations with $U=2.5$ eV.}
\begin{ruledtabular}
\begin{tabular}{ldddd}
   $P$ (GPa)  & \multicolumn{1}{r}{\textrm{$a$ ({\AA})}} & \multicolumn{1}{r}{\textrm{$b$ ({\AA})}} &  \multicolumn{1}{r}{\textrm{$c$ ({\AA})}} &  \multicolumn{1}{c}{\textrm{$V$ ({\AA}$^{3}$)}} \\
   \colrule
  0 & 4.7364 & 10.4252 & 6.0552 & 298.99 \\
 10 & 4.6013 & 10.0379 & 5.9122 & 273.07 \\
 20 & 4.5100 & 9.7447 & 5.8136 & 255.5\\
 30 & 4.4397 & 9.5046 & 5.7418 &  242.29\\
 40 & 4.3810 & 9.3005 &  5.6877 & 231.75\\
 50 & 4.3293 & 9.1236 &  5.6464 & 223.03 \\
 60 & 4.2824 & 8.9679 & 5.6146 & 215.62\\
 70 & 4.2388 & 8.8291 & 5.5897 & 209.19\\
 80 & 4.1980 & 8.7035 & 5.5703 & 203.52\\
 90 & 4.1601 & 8.5871 & 5.5555 & 198.46\\
100 & 4.1250 & 8.4799 & 5.5430 & 193.89\\
\end{tabular}
\end{ruledtabular}
\end{table*}

\begin{table*}%
\caption{\label{Table_S4}%
Lattice parameters and volume for the HP-$Pbnm$ LiFePO$_{4}$ phase in the LS state from our GGA+U calculations with $U=2.5$ eV.}
\begin{ruledtabular}
\begin{tabular}{ldddd}
   $P$ (GPa)  & \multicolumn{1}{r}{\textrm{$a$ ({\AA})}} & \multicolumn{1}{r}{\textrm{$b$ ({\AA})}} &  \multicolumn{1}{r}{\textrm{$c$ ({\AA})}} &  \multicolumn{1}{c}{\textrm{$V$ ({\AA}$^{3}$)}} \\
   \colrule
   0  & 4.6825 & 10.1545 & 5.9303 & 286.04 \\
   10 & 4.5842 & 9.7149 & 5.6419 & 251.27 \\
   20 & 4.5127 & 9.5110 & 5.5236 & 237.07\\
   30 & 4.4565 & 9.3426 & 5.4318 &  226.16\\
   40 & 4.4105 & 9.1978 &  5.3565 & 217.29\\
   50 & 4.3711 & 9.0706 &  5.2935 & 209.88 \\
   60 & 4.3367 & 8.9575 & 5.2389 & 203.51\\
   70 & 4.3064 & 8.8547 & 5.1910 & 197.94\\
   80 & 4.2790 & 8.7615 & 5.1480 & 193.00\\
   90 & 4.2544 & 8.6757 & 5.1090 & 188.58\\
  100 & 4.2317 & 8.5966 & 5.0736 & 184.57\\
\end{tabular}
\end{ruledtabular}
\end{table*}

\begin{figure*}
\includegraphics[width=0.9\textwidth]{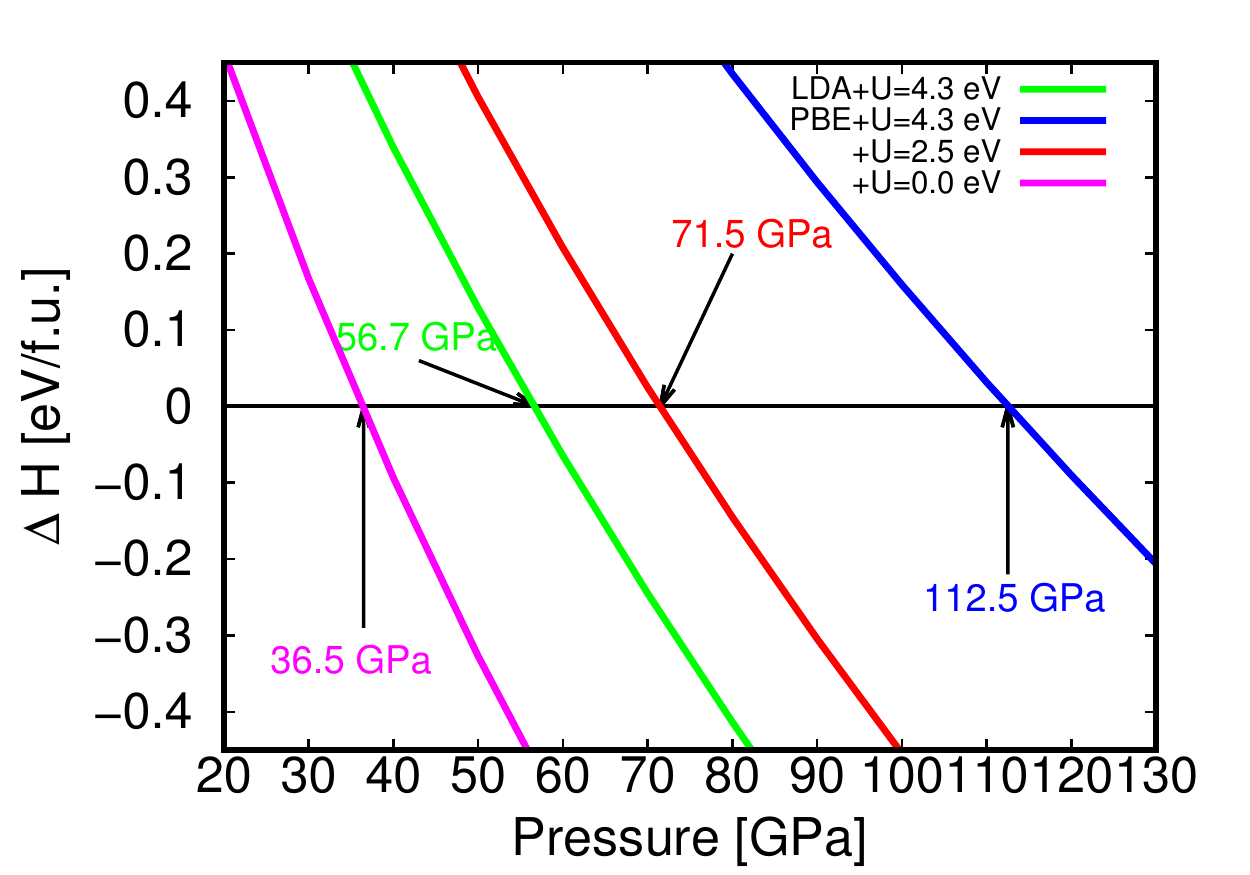}
\caption{\label{figS3} Relative enthalpies of $Pbnm-$LiFePO$_4$ in LS state (with the HS as reference) obtained using different on-site Coulomb interaction $U$ and exchange-correlation functional.}
\end{figure*}

\begin{figure*}
\includegraphics[width=0.9\textwidth]{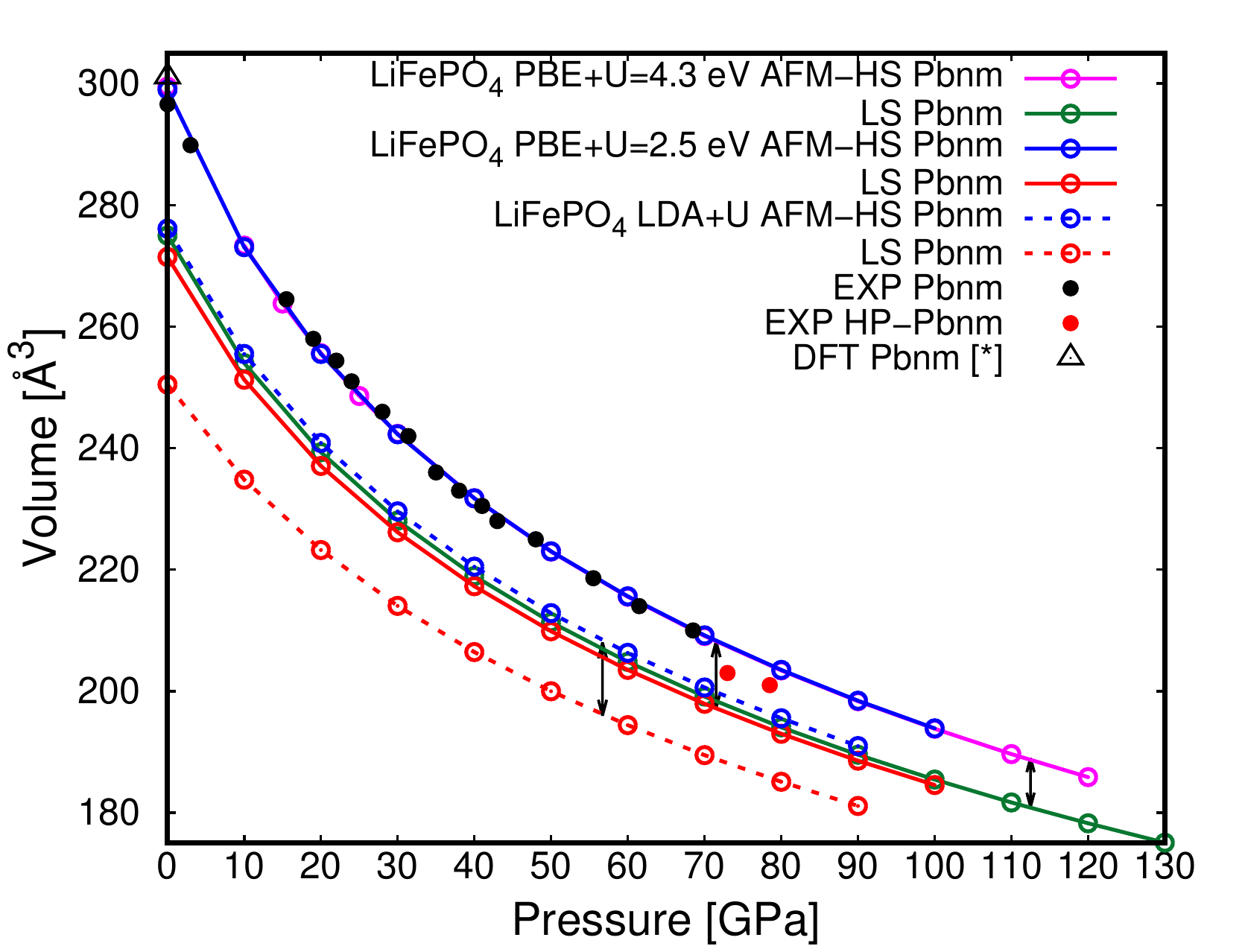}
\caption{\label{figS4} Comparison of our calculated volume using $U=2.5$ eV and $U=4.3$ eV within GGA+U, and $U=4.3$ eV within LDA+U. EXP data correspond to our measurements and DFT-point [$\ast$] is from Ref. [\onlinecite{Zhou2004}]. Double-head arrows indicate the difference in volume between the HS and the LS states at the transition pressures predicted for the corresponding $U$ value (See Fig. 3). We found that $U=2.5$ eV describes better both structural parameters and transition pressure in LiFeO$_4$ when contrasted with our experimental results.}
\end{figure*}
\bibliography{MariNV_library}